\documentclass[12pt,letterpaper]{article}

\usepackage[left=1in,right=1in,top=1in,bottom=1in]{geometry}
\usepackage{setspace}

\usepackage{amsmath,amssymb,amsthm,mathtools,bm}

\usepackage{graphicx}
\usepackage{booktabs}
\usepackage{threeparttable}
\usepackage{longtable}
\usepackage{array}
\usepackage{multirow}
\usepackage{caption}
\usepackage{subcaption}
\usepackage{float}
\usepackage{placeins}
\usepackage{makecell}

\usepackage{enumitem}
\usepackage{xcolor}
\usepackage{appendix}
\usepackage{pdflscape}
\usepackage{csquotes}

\usepackage{chngcntr}
\counterwithin{figure}{section}
\counterwithin{table}{section}

\usepackage[authoryear]{natbib}

\usepackage[colorlinks=true,linkcolor=blue,citecolor=blue,urlcolor=blue]{hyperref}
\usepackage[nameinlink,capitalize]{cleveref}

\theoremstyle{definition}

\title{Tuning in to Frequencies:\\How Global Assets Align with\\U.S. Put--Call Parity Residuals}
\author{Useong Shin\thanks{
		Sogang Business School, Sogang University (Seoul, Korea).\\
		ORCID: \href{https://orcid.org/0009-0003-0197-9003}{0009-0003-0197-9003}\\
		Email: \texttt{useong@sogang.ac.kr}
}}
\date{\today}

\begin{document}
	
	\maketitle
	\thispagestyle{empty}
	
	\begin{flushleft}
	\end{flushleft}
	
	\noindent\textbf{Acknowledgments:}
	I am grateful to Michele Azzone (Politecnico di Milano) for generously sharing OIS data, for guidance on implementing the implied-discount-factor pipeline, and for detailed feedback on earlier drafts. All remaining errors are my own.
	
	\begin{abstract}
		Put--call parity is risk-neutral at terminal payoff, but its enforcement is path-dependent and capital-using. I test whether the SPX and RUT carry gap is explained by OIS-based funding, volatility, trading-friction, and financial-condition variables, or also by residual outside-option information. Adding IEFA, IGOV, and IAU improves in-sample and leave-one-year-out fit after U.S.-centered controls. Gains survive broad-dollar neutralization, alternative blocks, PCA, residualization, and nested horizon selection. Results support reduced-form P--Q alignment: finite-capital parity enforcement reflects physical-measure investment opportunities, not payoff-level no-arbitrage failure.
	\end{abstract}
	
	\pagenumbering{arabic}
	
	\newpage
	\section{Introduction}
	\label{sec:intro}
	
	Put--call parity in European index options is among the cleanest no-arbitrage
	relations observable in real markets. Early exercise is absent, contracts are
	standardized, and exchange-traded futures provide a direct instrument for the
	forward leg. If any market should compress visible parity deviations, SPX and RUT
	index options should.
	
	This clean setting makes the remaining wedge more informative. The issue is not
	that quoted put--call parity visibly fails. Rather, even when the price-space
	residual is tightly compressed, the option-implied discount factor can differ
	systematically from the OIS benchmark. I study this annualized difference as the
	carry gap: a carry-space residual associated with the implementation of parity
	rather than a violation of the terminal-payoff identity.
	
	This paper builds on two earlier steps in the same framework. \citet{Shin26a}
	shows that the carry gap is related to a GBM path-risk term of the form
	\(r\sigma\sqrt{\tau}\), consistent with the idea that parity enforcement is
	capital-using before maturity. \citet{Shin26b} shows that a prior physical-drift
	proxy, \(\hat{\mu}\), adds explanatory power through an \(r\hat{\mu}\tau\) term.
	Together, these results suggest that the enforcement of a risk-neutral parity
	relation can be exposed to physical-measure states through the implementation
	path, even though the terminal payoff identity itself remains risk-neutral.
	
	The present paper asks whether this physical-measure component is better viewed
	as a narrow own-index drift effect or as part of a broader outside-option state.
	If capital is tied to parity enforcement, it cannot be deployed elsewhere. The
	relevant opportunity cost may therefore be only partly summarized by OIS rates.
	It may also reflect low-frequency investment opportunities available to the
	finite-capital arbitrageurs who enforce parity. I test this idea by adding a
	parsimonious 3ETF block---IEFA, IGOV, and IAU---to the OIS-based baseline. These
	assets proxy for developed ex-U.S. equity, international sovereign bonds, and
	gold, respectively. Each component is constructed from a prior rolling OLS slope
	of the log price path, using only information available through \(t-1\).
	
	The 3ETF block substantially improves the baseline in both markets. In-sample
	\(R^2\) rises from 0.312 to 0.402 in SPX and from 0.281 to 0.361 in RUT.
	Leave-one-year-out pooled out-of-sample \(R^2\) rises from 0.221 to 0.379 in SPX
	and from 0.171 to 0.318 in RUT. The result is not confined to a single holdout
	year and survives broad-dollar neutralization, PCA and residualization tests,
	alternative asset blocks, and nested horizon selection. The improvement is
	therefore difficult to interpret as a simple in-sample fit gain or a disguised
	U.S.-dollar factor.
	
	The absorption pattern is also informative. Once the 3ETF block is included, the
	own-index \(\hat{\mu}\) proxy adds little incremental explanatory power. Thus,
	the drift proxy is not rejected; rather, it appears to be a scalar projection of
	a broader physical-measure investment-opportunity state. The evidence should be
	read as reduced-form P--Q alignment, not as a structural P-to-Q transmission
	model. Put--call parity remains a terminal-payoff identity. The claim is narrower:
	a carry-space residual constructed from option-implied and OIS discount factors
	can align with physical-measure outside-option components because the
	risk-neutral parity relation is enforced by finite-capital intermediaries.
	
	The rest of the paper proceeds as follows. Section~\ref{sec:lit} reviews the
	related literature. Section~\ref{sec:motivation} motivates the outside-option
	channel. Section~\ref{sec:data} describes the carry gap, the baseline
	specification, and the asset-return-based GBM terms. Section~\ref{sec:3ETF}
	derives the 3ETF specification. Section~\ref{sec:results} reports the main
	empirical results. Section~\ref{sec:horse_race} compares the 3ETF specification
	with the drift-extended specification. Section~\ref{sec:robust} presents
	robustness checks. Section~\ref{sec:discuss} discusses interpretation and
	limitations. Section~\ref{sec:conclusion} concludes.
	Appendix~\ref{app:implementation} provides data-processing and implementation
	details.
	
	\section{Related Literature}
	\label{sec:lit}
	
	This paper connects three literatures: put--call parity and option-implied
	discounting, limits to arbitrage and intermediary capital, and the relation
	between risk-neutral objects and physical-measure state variables. It also
	directly extends \citet{Shin26a} and \citet{Shin26b}. The common theme is that
	no-arbitrage relations are exact at the terminal-payoff level, but their
	enforcement in actual markets can be path-dependent, capital-using, and exposed
	to physical-measure investment opportunities.
	
	\subsection{Put--Call Parity and Implied Discounting}
	\label{subsec:lit_pcp_discount}
	
	Put--call parity has been a basic no-arbitrage relation in option pricing since
	\citet{Stoll69}. The empirical literature asks whether observed parity
	deviations represent genuine arbitrage opportunities or instead reflect
	transaction costs, short-sale constraints, execution frictions, and market
	microstructure noise \citep{GG74,KR79,AT01}. The common implication is that
	observed price-space deviations should not be interpreted mechanically as
	risk-free arbitrage opportunities.
	
	I study a different object. Using the option-implied discount-factor procedure
	of \citet{AB21}, I estimate discount factors from SPX and RUT option
	cross-sections and compare them with OIS discount factors. This yields the carry
	gap, an annualized carry-space wedge. The question is therefore not whether
	quoted put--call parity visibly fails, but which state variables align with the
	wedge that remains after visible parity deviations have largely been compressed.
	
	\subsection{Limits to Arbitrage after Visible Parity Is Closed}
	\label{subsec:lit_limits_intermediary}
	
	The interpretation builds on the limits-to-arbitrage literature. Since
	\citet{SV97}, this literature has emphasized that arbitrage is performed by
	finite-capital intermediaries who face funding constraints, margin requirements,
	interim losses, and liquidation risk \citep{GV02,BP09,MP12}. In options markets,
	\citet{ORW04} show that short-sale constraints and limits to arbitrage can be
	linked to put--call parity deviations.
	
	My setting is deliberately different from the standard short-sale-constraint
	environment. Much of the existing evidence studies cases in which visible
	price-space deviations remain open because arbitrage is directly impeded. Here,
	the visible put--call parity residual is already tightly compressed in liquid
	European index-option markets. The question is why a systematic carry-space
	wedge remains after the obvious price-space arbitrage has largely been closed.
	
	This view is also consistent with work linking intermediary capital to
	law-of-one-price deviations \citep{GP11,HK13,AEM14,HKM17}. It is close in spirit
	to the CIP evidence in \citet{DTV18}: enforcing no-arbitrage relations uses
	balance-sheet capacity, capital, and margin, and the marginal cost of that
	enforcement varies with market conditions. The contribution here is to apply
	this logic to a clean parity setting and to ask whether the remaining
	carry-space wedge is aligned with outside investment opportunities faced by
	finite-capital arbitrageurs.
	
	\subsection{Risk-Neutral Objects and Physical-Measure States}
	\label{subsec:lit_pq_state}
	
	This paper also relates to work connecting option-implied, risk-neutral objects
	to physical-measure information. \citet{BTZ09}, \citet{Ross15}, and
	\citet{Martin17} show that option-implied \(Q\)-measure objects can contain
	information about variance risk premia, physical probabilities, or expected
	returns.
	
	My direction is complementary. I do not recover physical probabilities from
	option prices, and I do not identify a structural mapping from physical returns
	to risk-neutral prices. Instead, I ask whether a \(Q\)-measure carry-space
	residual is empirically aligned with \(P\)-measure outside-option proxies because
	the parity relation is enforced by finite-capital intermediaries. This question
	is related to investment-opportunity-state models such as \citet{Merton73}'s
	ICAPM, \citet{Ross76}'s APT, and multi-factor asset-pricing frameworks
	\citep{Campbell93,CV04,CRR86,FF93,Petkova06}. The exercise is not a structural
	ICAPM or APT test; it is a reduced-form test of whether the opportunity cost of
	parity-enforcement capital is aligned with multiple physical-measure state
	variables.
	
	\subsection{Direct Antecedents}
	\label{subsec:lit_direct}
	
	This paper directly extends \citet{Shin26a} and \citet{Shin26b}. \citet{Shin26a}
	documents the carry gap and shows that OIS-based GBM path-risk terms, trading
	frictions, and financial conditions explain a meaningful part of its variation.
	The key reduced-form object is an \(r\sigma\sqrt{\tau}\) implementation-risk
	term.
	
	\citet{Shin26b} extends the framework by preserving physical drift in the
	support-capital logic. Since the true physical drift \(\mu\) is unobserved, that
	paper uses a prior rolling trend in the own-index total return as a proxy
	\(\hat{\mu}\), and shows that an \(r\hat{\mu}\tau\) term adds explanatory power
	for the carry gap. The present paper asks whether this own-index drift proxy is
	an independent state variable or a scalar projection of a broader
	outside-option state. To test this, I introduce a 3ETF block consisting of IEFA,
	IGOV, and IAU, and compare the baseline, drift, 3ETF, and 3ETF+drift
	specifications.
	
	\section{Motivation for the Asset Extension}
	\label{sec:motivation}
	
	This section motivates the extension from the OIS-based baseline to
	asset-return components. The key idea is simple. Capital tied to parity
	enforcement cannot be deployed elsewhere. Even if the terminal payoff of
	put--call parity is fixed, the enforcing strategy must be carried through
	variation margin, funding needs, trading frictions, and finite-capital
	constraints. The carry gap may therefore respond not only to OIS rates, but also
	to the outside investment opportunities forgone by arbitrageurs.
	
	Consider a finite-capital arbitrageur, Bob. Bob observes a small put--call parity
	wedge in the SPX or RUT options market and enters a parity-enforcement position
	combining a synthetic forward and a futures leg. At the terminal-payoff level,
	the trade is nearly locked. Before maturity, however, Bob must absorb daily
	mark-to-market gains and losses and meet variation-margin calls. If supporting
	the trade requires reducing other risky or liquid positions, Bob forgoes the
	expected return, liquidity service, or balance-sheet capacity those positions
	would have provided. Thus, even when the terminal payoff is fixed, the
	implementation path has an opportunity cost.
	
	From this perspective, restricting the opportunity-cost component of the
	baseline GBM term to OIS rates may be too narrow. OIS rates are a natural
	benchmark for risk-free funding costs, but the actual capital-allocation problem
	of a finite-capital arbitrageur is broader. Bob compares harvesting the parity
	wedge with deploying the same capital elsewhere. The shadow cost of
	parity-enforcement capital may therefore reflect low-frequency outside-option
	states as well as risk-free rates.
	
	The asset extension implements this idea inside the GBM structure. The baseline
	specification can be written conceptually as
	\[
	\text{GBM term}
	=
	\text{path-risk scale}
	\times
	\text{opportunity-cost component}.
	\]
	In the baseline, the opportunity-cost component is proxied by OIS rates. In the
	3ETF extension, I expand this component using low-frequency return components of
	global equity, sovereign bonds, and gold. The 3ETF block is therefore not a
	generic set of added predictors. It is a reduced-form proxy for the outside-option
	set faced by a finite-capital arbitrageur.
	
	This interpretation also connects to the drift extension. \citet{Shin26b} shows
	that an own-index rolling drift proxy, \(\hat{\mu}\), adds explanatory power for
	the carry gap through an \(r\hat{\mu}\tau\) term. I ask whether this proxy is an
	independent scalar primitive, or instead a compressed proxy for a broader
	physical-measure state that partly overlaps with U.S. financial conditions and
	partly extends beyond them. To examine this, I compare the baseline, drift,
	3ETF, and 3ETF+drift specifications.
	
	The empirical predictions are direct. If the carry gap is an OIS-contained wedge,
	adding asset-return components should produce only limited gains over the
	baseline. If the carry gap is aligned with outside-option states, low-frequency
	components from equities, bonds, and gold should add explanatory power. If the
	3ETF block absorbs the information in the own-index \(\hat{\mu}\) proxy, then
	the drift effect is better interpreted as a scalar projection of a broader
	residual opportunity-cost state rather than as an independent primitive.
	
	\section{Data and Methodology}
	\label{sec:data}
	
	This section summarizes the carry-gap measure, the OIS-based baseline,
	the asset-return GBM terms, and the comparison models. Appendix~\ref{app:implementation}
	provides implementation details on data cleaning, option-implied discount-factor
	identification, ETF slope construction, broad-dollar adjustment, HAC inference,
	and leave-one-year-out evaluation.
	
	\subsection{Carry Gap}
	
	I study European-style SPX and RUT index options using minute-level NBBO quotes
	from ThetaData. The sample runs from January 4, 2016 to October 31, 2025, the
	period over which the OIS data used in the baseline are available. Because both
	markets use European-style index options, early-exercise premia do not directly
	affect the put--call-parity-based identification of discount factors.
	
	Following \citet{AB21}, I estimate the option-implied discount factor
	\(\hat B_t(T)\) from synthetic-forward relations and compare it with the OIS
	discount factor \(D_t^{\mathrm{OIS}}(T)\). Let \(\tau_t(T)=T-t\). The carry gap is
	\begin{equation}
		CG_t(T)
		=
		\frac{1}{\tau_t(T)}
		\log\!\left(
		\frac{D_t^{\mathrm{OIS}}(T)}{\hat B_t(T)}
		\right),
		\label{eq:carry_gap_def}
	\end{equation}
	and I use the basis-point version
	\begin{equation}
		CG_t^{bp}(T)=10^4\cdot CG_t(T).
		\label{eq:carry_gap_bp}
	\end{equation}
	The regression outcome, \(CG_{i,t}^{bp}\), is the cleaned daily market-level
	carry gap for market \(i\in\{\mathrm{SPX},\mathrm{RUT}\}\).
	
	\subsection{Baseline and Asset-Return GBM Terms}
	
	The baseline is an OIS-based GBM path-risk specification:
	\begin{equation}
		CG_{i,t}^{bp}
		=
		\alpha_i
		+\phi_{1,i}GBM_{i,t}^{\mathrm{OIS},1Y}
		+\phi_{10,i}GBM_{i,t}^{\mathrm{OIS},10Y}
		+\beta_i\frac{BA_{i,t}^{med}}{\tau_{i,t}}
		+\gamma_iNFCI_t
		+\varepsilon_{i,t}.
		\label{eq:baseline_reg}
	\end{equation}
	The two OIS GBM terms use the one-year and ten-year OIS rates as
	opportunity-cost components. The variable \(BA_{i,t}^{med}/\tau_{i,t}\) captures
	ATM option-market trading frictions, and \(NFCI_t\) is the Chicago Fed National
	Financial Conditions Index.
	
	Conceptually,
	\[
	\text{GBM term}
	=
	\text{path-risk scale}
	\times
	\text{opportunity-cost component}.
	\]
	The asset extension keeps this structure but replaces part of the
	opportunity-cost component with low-frequency outside-option return proxies. For
	asset \(a\) and lookback window \(n\), define
	\begin{equation}
		GBM_{i,t}^{a,n}
		=
		10^{4}
		\cdot
		b_{a,t}^{(n)}
		\cdot
		\frac{2}{3}
		\cdot
		\frac{Vol_{i,t}}{100}
		\cdot
		\sqrt{\frac{2\tau_{i,t}}{\pi}},
		\label{eq:gbm_term_asset}
	\end{equation}
	where \(b_{a,t}^{(n)}\) is the prior \(n\)-day rolling OLS slope of ETF \(a\)'s
	log price path, computed using information through \(t-1\). The volatility input
	\(Vol_{i,t}\) is VIX for SPX and RVX for RUT. Thus, the asset-return term places
	an outside-option return proxy in the slot occupied by the OIS rate in the
	baseline GBM term.
	
	\subsection{Comparison Models and Evaluation}
	
	I compare four specifications. The baseline includes the OIS 1Y and OIS 10Y GBM
	terms. The drift specification adds the own-index rolling drift proxy
	\(\hat{\mu}\), since the true physical drift \(\mu\) is unobserved. The 3ETF
	specification keeps OIS 1Y, drops OIS 10Y, and adds IEFA, IGOV, and IAU GBM
	terms. The 3ETF+drift specification adds the own-index \(\hat{\mu}\) proxy to
	the 3ETF model.
	
	The comparison asks two questions. First, do outside-option components improve
	on the OIS-only baseline? Second, does the own-index \(\hat{\mu}\) proxy retain
	independent explanatory power once the 3ETF block is included? If 3ETF+drift
	adds little to 3ETF, the drift proxy is better interpreted as a scalar projection
	of the broader residual opportunity-cost state.
	
	All regressions are estimated separately for SPX and RUT. Coefficient inference
	uses date-based HAC (Newey--West) standard errors. Out-of-sample performance is
	evaluated by leave-one-year-out validation.
	
	\section{Selecting the 3ETF Block}
	\label{sec:3ETF}
	
	This section explains how I select the 3ETF specification. The exercise is a
	restricted asset-allocation scan, not an unrestricted predictor search. I begin
	with standard outside-option categories---equities, bonds, and gold---and vary
	the regional equity and bond components across U.S., developed ex-U.S., and
	emerging-market blocks.
	
	\subsection{Candidate Assets}
	\label{subsec:3etf_candidates}
	
	I first compute the incremental \(R^2\) over the OIS-based baseline for each
	candidate ETF across lookback horizons. The candidate set contains ten ETFs
	covering major asset-allocation categories.\footnote{
		VTI (U.S. equity), IEFA (developed ex-U.S. equity),
		IEMG (emerging-market equity), BND (U.S. aggregate bond),
		SCHP (U.S. inflation-linked bond), IGOV (developed ex-U.S. sovereign
		bond, FX-unhedged), EBND (emerging-market sovereign bond, FX-unhedged),
		IAU (gold), VNQ (U.S. REITs), and VNQI (ex-U.S. REITs).}
	For each ETF and lookback window, I convert the prior log-price slope into the
	asset-return GBM term in Section~\ref{sec:data} and measure the fit gain
	relative to the baseline.
	
	\FloatBarrier
	\begin{figure}[H]
		\centering
		\includegraphics[width=6.5in]{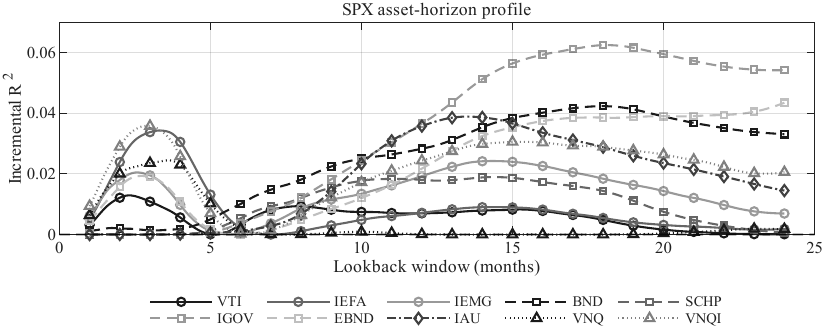}
		\includegraphics[width=6.5in]{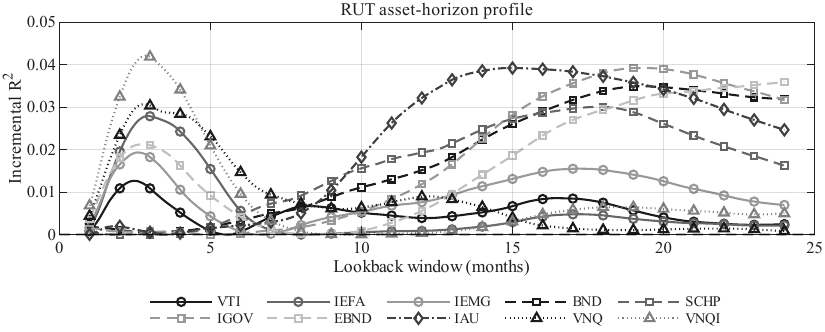}
		\caption{Incremental \(R^2\) by lookback horizon for ETFs representing major asset classes.
			The upper panel reports SPX results; the lower panel reports RUT results.}
		\label{fig:all_horizon}
	\end{figure}
	\FloatBarrier
	
	Figure~\ref{fig:all_horizon} shows that the useful outside-option information is
	not concentrated at a single frequency. IEFA is informative at relatively short
	horizons, IGOV at much longer horizons, and IAU at medium-to-long horizons. The
	peaks are also surrounded by broad high-\(R^2\) plateaus, so the selected
	lookbacks should be read as representative horizon bands rather than pointwise
	optimized constants.
	
	\subsection{Selected Horizons}
	\label{subsec:3etf_selected}
	
	Following the asset-allocation taxonomy, I compare three blocks: a U.S.-centered
	block \((\mathrm{VTI},\mathrm{BND},\mathrm{IAU})\), a developed ex-U.S. block
	\((\mathrm{IEFA},\mathrm{IGOV},\mathrm{IAU})\), and an emerging-market block
	\((\mathrm{IEMG},\mathrm{EBND},\mathrm{IAU})\). All three improve on the
	OIS-only baseline. The developed ex-U.S. block performs best overall, so I use
	IEFA, IGOV, and IAU as the main 3ETF specification. Section~\ref{sec:robust}
	revisits the other two blocks as alternative-asset robustness checks.
	
	The final lookback windows are 70 trading days for IEFA, 441 trading days for
	IGOV, and 315 trading days for IAU. IEFA represents a fast developed ex-U.S.
	equity component, IGOV a slow international sovereign-bond component, and IAU an
	intermediate gold component.
	
	\FloatBarrier
	\begin{figure}[H]
		\centering
		\includegraphics[width=6.5in]{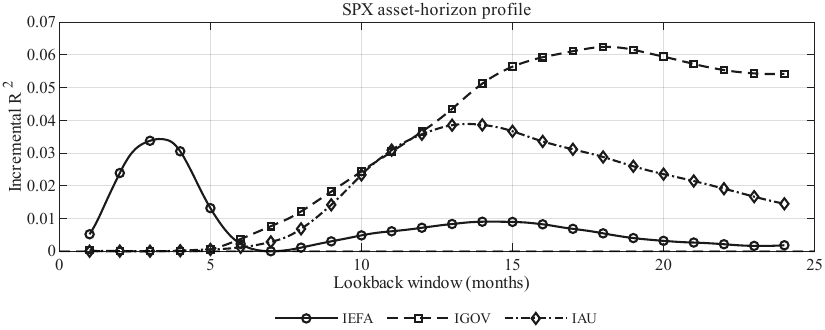}\\[0.5em]
		\includegraphics[width=6.5in]{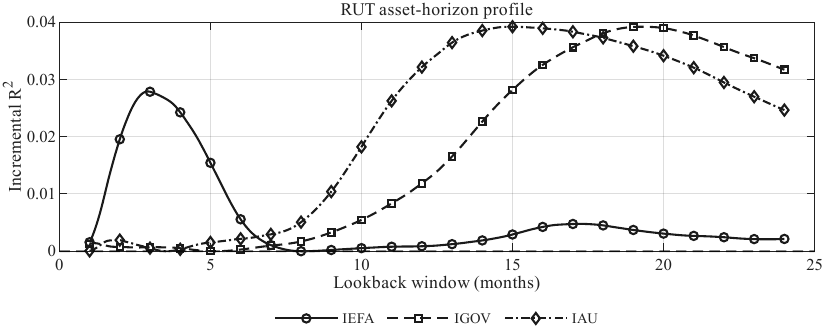}
		\caption{Incremental \(R^2\) by lookback horizon for the three selected ETFs.
			The upper panel reports SPX results; the lower panel reports RUT results.}
		\label{fig:3_spx_rut}
	\end{figure}
	\FloatBarrier
	
	Figure~\ref{fig:3_spx_rut} isolates the three retained components. The selected
	ETFs line up with different horizon bands in both SPX and RUT: IEFA at short
	horizons, IGOV at long horizons, and IAU between them. This frequency
	heterogeneity is the reason the final block uses three assets rather than a
	single outside-option proxy.
	
	\begin{figure}[H]
		\centering
		\includegraphics[width=6.5in]{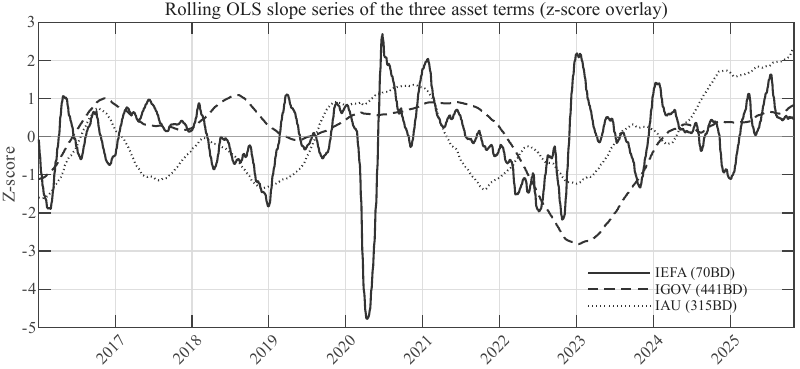}
		\caption{OLS slope series of the three selected ETFs.
			Each series is the log OLS slope under the lookback window used in the main specification
			(IEFA: 70 days, IGOV: 441 days, IAU: 315 days),
			standardized as a \(z\)-score.}
		\label{fig:olsseries}
	\end{figure}
	
	Figure~\ref{fig:olsseries} shows the same point in the time-series domain. IEFA
	moves relatively quickly, IGOV moves slowly, and IAU lies between them. The 3ETF
	block therefore combines outside-option components with distinct low-frequency
	speeds rather than stacking parallel versions of the same signal. Section~\ref{sec:robust}
	tests this interpretation directly using PCA and residualization.
	
	\subsection{Final Specification}
	\label{subsec:3etf_final}
	
	The final 3ETF regression keeps the OIS 1Y GBM term, drops the OIS 10Y GBM term,
	and adds the IEFA, IGOV, and IAU GBM terms. The exclusion of OIS 10Y is
	empirical: IGOV overlaps with the long-horizon opportunity-cost variation
	previously captured by OIS 10Y, and including both terms destabilizes the
	coefficient structure with little gain in fit. The selected block therefore
	replaces the long-horizon OIS component rather than being stacked mechanically
	on top of the full baseline.
	
	For market \(i\in\{\mathrm{SPX},\mathrm{RUT}\}\), the final regression is
	\begin{equation}
		\begin{aligned}
			CG_{i,t}^{bp}
			={}&\alpha_i
			+\phi_{1,i}GBM_{i,t}^{\mathrm{OIS},1Y} \\
			&+\theta_{E,i}GBM_{i,t}^{\mathrm{IEFA},70}
			+\theta_{G,i}GBM_{i,t}^{\mathrm{IGOV},441}
			+\theta_{A,i}GBM_{i,t}^{\mathrm{IAU},315} \\
			&+\beta_i\frac{BA_{i,t}^{med}}{\tau_{i,t}}
			+\gamma_iNFCI_t
			+\varepsilon_{i,t}.
		\end{aligned}
		\label{eq:final_3etf_reg}
	\end{equation}
	
	The specification layers a restricted outside-option block on top of the
	short-horizon OIS opportunity-cost component. The next section evaluates how much
	this structure improves in-sample and out-of-sample fit relative to the
	OIS-based baseline.

	\section{Empirical Results}
	\label{sec:results}
	
	This section compares the OIS-based baseline with the 3ETF asset-return
	extension. The baseline includes OIS 1Y and OIS 10Y GBM path-risk terms. The
	3ETF specification keeps OIS 1Y, drops OIS 10Y, and adds IEFA, IGOV, and IAU
	GBM terms. The central question is whether the outside-option block improves fit
	even after the long-horizon OIS component is removed.
	
	\subsection{In-Sample Performance}
	\label{subsec:is_performance}
	
	\FloatBarrier
	\begin{table}[H]
		\centering
		\onehalfspacing
		\footnotesize
		\caption{In-sample performance: baseline versus 3ETF specification}
		\label{tab:main_is_compare}
		\begin{tabular}{lccccc}
			\toprule
			Market & \(R^2\) (baseline) & \(R^2\) (3ETF) & \(\Delta R^2\) & \(\Delta\)RMSE (bp) & \(\Delta\)MAE (bp) \\
			\midrule
			SPX & 0.312 & 0.402 & 0.090 & -0.893 & -0.942 \\
			RUT & 0.281 & 0.361 & 0.080 & -0.804 & -0.687 \\
			\bottomrule
		\end{tabular}
	\end{table}
	\FloatBarrier
	
	Table~\ref{tab:main_is_compare} shows that the 3ETF specification improves the
	maturity-pooled fit in both markets. In SPX, \(R^2\) rises from 0.312 to 0.402.
	In RUT, it rises from 0.281 to 0.361. RMSE and MAE decline in both markets.
	
	The gain is not a mechanical consequence of stacking more predictors on the full
	baseline. The 3ETF specification removes the OIS 10Y GBM term and still improves
	fit. This suggests that part of the low-frequency carry-gap variation previously
	assigned to long-horizon OIS rates is better captured by residual asset-return
	components conditional on the U.S.-centered baseline.
	
	\FloatBarrier
	\begin{figure}[H]
		\centering
		\includegraphics[width=6.5in]{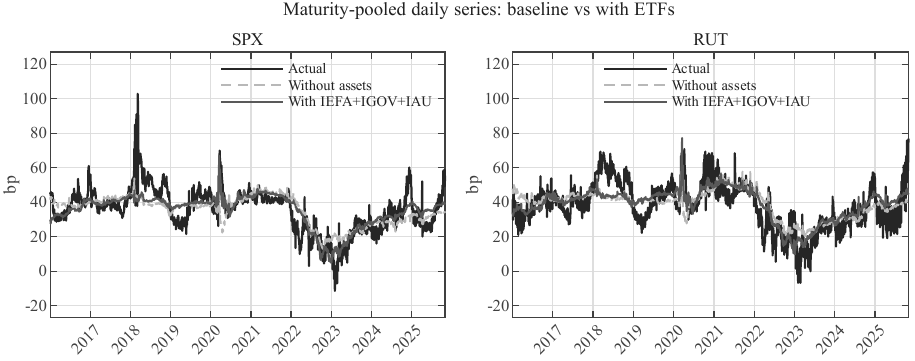}
		\caption{Maturity-pooled daily fit: baseline versus 3ETF specification}
		\label{fig:pooledfit}
	\end{figure}
	\FloatBarrier
	
	Figure~\ref{fig:pooledfit} gives the corresponding time-series view. In both
	markets, the 3ETF fitted value tracks low-frequency movements in the carry gap
	more closely than the baseline, especially around the post-2020 decline and the
	2024--2025 recovery. The improvement is therefore visible not only in summary
	fit statistics, but also along the daily carry-gap path.
	
	\subsection{Performance by Maturity Bin}
	\label{subsec:maturity_bin_performance}
	
	\FloatBarrier
	\begin{table}[H]
		\centering
		\onehalfspacing
		\footnotesize
		\caption{In-sample performance by maturity bin: baseline versus 3ETF specification}
		\label{tab:tau_bin_is_compare}
		\begin{tabular}{llrrrrr}
			\toprule
			Market & Maturity bin & \(R^2\) (baseline) & \(R^2\) (3ETF) & \(\Delta R^2\) & \(\Delta\)RMSE (bp) & \(\Delta\)MAE (bp) \\
			\midrule
			SPX & 1--2m   & 0.096 & 0.136 & 0.039 & -0.502 & -0.577 \\
			SPX & 2--3m   & 0.184 & 0.257 & 0.073 & -0.815 & -0.676 \\
			SPX & 3--5m   & 0.283 & 0.401 & 0.118 & -1.165 & -0.866 \\
			SPX & 5--7m   & 0.373 & 0.532 & 0.159 & -1.555 & -1.018 \\
			SPX & 7--10m  & 0.472 & 0.630 & 0.157 & -1.590 & -1.101 \\
			SPX & 10--14m & 0.525 & 0.621 & 0.096 & -0.884 & -0.905 \\
			SPX & 14--21m & 0.306 & 0.359 & 0.054 & -0.508 & -1.118 \\
			SPX & 21m+    & 0.182 & 0.195 & 0.013 & -0.106 & -1.209 \\
			\midrule
			RUT & 1--2m   & 0.112 & 0.164 & 0.052 & -0.665 & -0.537 \\
			RUT & 2--3m   & 0.205 & 0.304 & 0.099 & -1.007 & -0.633 \\
			RUT & 3--5m   & 0.254 & 0.370 & 0.116 & -1.047 & -0.761 \\
			RUT & 5--7m   & 0.231 & 0.429 & 0.198 & -1.737 & -1.352 \\
			RUT & 7--10m  & 0.284 & 0.461 & 0.177 & -1.573 & -1.149 \\
			RUT & 10--14m & 0.472 & 0.550 & 0.077 & -0.801 & -0.848 \\
			RUT & 14--21m & 0.482 & 0.517 & 0.035 & -0.347 & -0.359 \\
			RUT & 21m+    & 0.413 & 0.256 & -0.157 &  1.348 &  0.755 \\
			\bottomrule
		\end{tabular}
	\end{table}
	\FloatBarrier
	
	Table~\ref{tab:tau_bin_is_compare} shows that the 3ETF gain is strongest at
	intermediate maturities. In SPX, \(\Delta R^2\) reaches 0.159 in the 5--7 month
	bin and 0.157 in the 7--10 month bin. In RUT, the corresponding gains are 0.198
	and 0.177. The pattern is consistent with the interpretation that the
	outside-option channel is most visible where the enforcement path is long enough
	for carry and margin costs to accumulate, but not so long that additional
	slow-moving maturity-specific components dominate.
	
	The far long end is more mixed. SPX still improves slightly at 21m+, but RUT
	deteriorates in that bin. I therefore interpret the maturity-bin evidence as
	supporting an intermediate-horizon outside-option channel, not as showing that
	the 3ETF block uniformly dominates the baseline at every maturity.
	
	\subsection{Out-of-Sample Performance}
	\label{subsec:oos_performance}
	
	The out-of-sample exercise should be read as a stability test, not as a
	real-time forecasting experiment. The sample spans only nine years and ten
	months, which makes a conventional expanding-window design unattractive:
	early training windows would be short and regime-specific, especially for
	low-frequency asset components. Leave-one-year-out validation instead uses the
	sample symmetrically by allowing every calendar year to serve once as a holdout
	period. Because each holdout model is estimated on all remaining years, the
	exercise is not a trading forecast. Its purpose is narrower: to test whether the
	3ETF relation is concentrated in a small number of calendar years, or whether it
	survives the exclusion of any single year.
	
	\FloatBarrier
	\begin{table}[H]
		\centering
		\onehalfspacing
		\footnotesize
		\caption{LOYO out-of-sample performance: baseline versus 3ETF specification}
		\label{tab:loyo_oos_summary}
		\begin{tabular}{llrrrrr}
			\toprule
			Market & Specification & Mean \(R^2\) & Median \(R^2\) & Pooled \(R^2\) & Mean RMSE (bp) & Mean correlation \\
			\midrule
			SPX & Baseline & 0.059 & 0.130 & 0.221 & 13.95 & 0.205 \\
			SPX & 3ETF     & 0.288 & 0.195 & 0.379 & 12.37 & 0.373 \\
			\midrule
			RUT & Baseline & 0.075 & 0.108 & 0.171 & 15.07 & 0.243 \\
			RUT & 3ETF     & 0.237 & 0.211 & 0.318 & 13.83 & 0.356 \\
			\bottomrule
		\end{tabular}
	\end{table}
	\FloatBarrier
	
	Table~\ref{tab:loyo_oos_summary} reports the LOYO summary. The 3ETF
	specification outperforms the baseline in both markets. In SPX, pooled OOS
	\(R^2\) rises from 0.221 to 0.379. In RUT, pooled OOS \(R^2\) rises from 0.171
	to 0.318. Mean OOS \(R^2\) and mean correlation also increase, while mean RMSE
	declines in both markets. The improvement therefore is not limited to in-sample
	fit; the asset-return block continues to explain carry-gap variation when each
	calendar year is excluded from estimation in turn.
	
	\FloatBarrier
	\begin{figure}[H]
		\centering
		\includegraphics[width=6.5in]{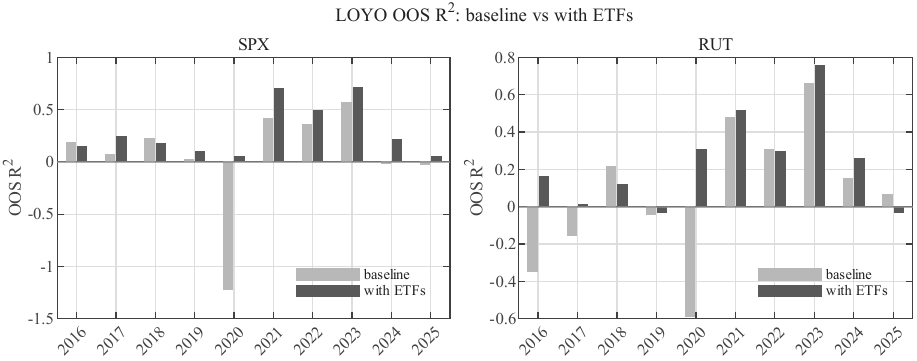}
		\caption{Year-by-year LOYO out-of-sample performance: baseline versus 3ETF specification}
		\label{fig:oos}
	\end{figure}
	\FloatBarrier
	
	Figure~\ref{fig:oos} compares OOS \(R^2\) by holdout year. The improvement is
	not driven by a single calendar year. The 3ETF model beats the baseline in
	8 out of 10 SPX holdouts and 7 out of 10 RUT holdouts. The largest repair occurs
	in 2020, when the baseline fit breaks down sharply. In SPX, OOS \(R^2\) moves
	from \(-1.221\) under the baseline to 0.050 under the 3ETF specification. In
	RUT, it moves from \(-0.587\) to 0.308. Thus, the 3ETF block helps most when the
	OIS-only baseline loses its level calibration.
	
	\FloatBarrier
	\begin{table}[H]
		\centering
		\onehalfspacing
		\footnotesize
		\caption{Year-by-year LOYO out-of-sample performance: SPX}
		\label{tab:oos_yearly_spx}
		\begin{tabular}{ccccccc}
			\toprule
			Holdout year &
			\multicolumn{3}{c}{Baseline} &
			\multicolumn{3}{c}{3ETF} \\
			\cmidrule(lr){2-4}\cmidrule(lr){5-7}
			& \(R^2\) & RMSE (bp) & Corr. & \(R^2\) & RMSE (bp) & Corr. \\
			\midrule
			2016 &  0.185 &  8.970 &  0.194 &  0.148 &  9.173 &  0.363 \\
			2017 &  0.074 & 10.069 &  0.226 &  0.238 &  9.133 &  0.394 \\
			2018 &  0.221 & 26.304 &  0.259 &  0.179 & 26.997 &  0.181 \\
			2019 &  0.023 &  8.934 & -0.011 &  0.094 &  8.604 &  0.271 \\
			2020 & -1.221 & 18.423 & -0.294 &  0.050 & 12.048 &  0.416 \\
			2021 &  0.416 &  7.600 &  0.064 &  0.701 &  5.436 &  0.092 \\
			2022 &  0.357 & 16.040 &  0.347 &  0.493 & 14.238 &  0.285 \\
			2023 &  0.571 & 15.725 &  0.438 &  0.710 & 12.932 &  0.509 \\
			2024 & -0.016 & 15.428 &  0.589 &  0.211 & 13.599 &  0.751 \\
			2025 & -0.022 & 11.980 &  0.242 &  0.053 & 11.529 &  0.473 \\
			\midrule
			Mean   &  0.059 & 13.947 & 0.205 &  0.288 & 12.369 & 0.373 \\
			Median &  0.130 &        &       &  0.195 &        &       \\
			Pooled &  0.221 &        &       &  0.379 &        &       \\
			\bottomrule
		\end{tabular}
	\end{table}
	\FloatBarrier
	
	\FloatBarrier
	\begin{table}[H]
		\centering
		\onehalfspacing
		\footnotesize
		\caption{Year-by-year LOYO out-of-sample performance: RUT}
		\label{tab:oos_yearly_rut}
		\begin{tabular}{ccccccc}
			\toprule
			Holdout year &
			\multicolumn{3}{c}{Baseline} &
			\multicolumn{3}{c}{3ETF} \\
			\cmidrule(lr){2-4}\cmidrule(lr){5-7}
			& \(R^2\) & RMSE (bp) & Corr. & \(R^2\) & RMSE (bp) & Corr. \\
			\midrule
			2016 & -0.350 & 12.787 &  0.069 &  0.164 & 10.066 &  0.267 \\
			2017 & -0.152 & 10.498 &  0.142 &  0.014 &  9.712 &  0.361 \\
			2018 &  0.213 & 18.830 &  0.121 &  0.120 & 19.911 &  0.203 \\
			2019 & -0.041 & 11.088 & -0.007 & -0.029 & 11.025 &  0.422 \\
			2020 & -0.587 & 22.437 &  0.102 &  0.308 & 14.809 &  0.437 \\
			2021 &  0.477 & 12.368 &  0.431 &  0.514 & 11.930 &  0.353 \\
			2022 &  0.308 & 16.619 &  0.368 &  0.295 & 16.776 &  0.238 \\
			2023 &  0.661 & 13.922 &  0.354 &  0.756 & 11.810 &  0.458 \\
			2024 &  0.153 & 14.005 &  0.598 &  0.258 & 13.110 &  0.486 \\
			2025 &  0.064 & 18.193 &  0.248 & -0.032 & 19.101 &  0.335 \\
			\midrule
			Mean   &  0.075 & 15.075 & 0.243 &  0.237 & 13.825 & 0.356 \\
			Median &  0.108 &        &       &  0.211 &        &       \\
			Pooled &  0.171 &        &       &  0.318 &        &       \\
			\bottomrule
		\end{tabular}
	\end{table}
	\FloatBarrier
	
	Tables~\ref{tab:oos_yearly_spx} and~\ref{tab:oos_yearly_rut} report the
	year-by-year results. In SPX, the 3ETF specification has a higher OOS \(R^2\)
	in all holdout years except 2016 and 2018. In RUT, it outperforms the baseline
	except in 2018, 2022, and 2025. In 2019, both RUT specifications have negative
	OOS \(R^2\), but the loss is smaller under the 3ETF specification.
	
	The main lesson is that the 3ETF block improves average fit and also mitigates
	some of the baseline's worst failures. The evidence is not uniform across every
	year: some holdouts still favor the baseline, and RUT 2025 remains negative
	under the 3ETF specification. I therefore do not interpret the 3ETF model as a
	complete forecasting model. The more appropriate interpretation is that the
	3ETF block captures a low-frequency outside-option component missed by the
	OIS-only baseline, improving overall out-of-sample stability and stress-year
	calibration.
	
	\subsection{Coefficient Structure}
	\label{subsec:coef_structure}
	
	This subsection examines the coefficient structure behind the fit gains. The
	full-sample HAC estimates show signs, magnitudes, and statistical significance.
	The LOYO training estimates show whether the coefficients used in the
	out-of-sample exercise are stable across holdout folds.
	
	\FloatBarrier
	\begin{table}[H]
		\centering
		\onehalfspacing
		\footnotesize
		\caption{In-sample coefficients: baseline versus 3ETF specification, HAC inference}
		\label{tab:coef_is_compare}
		\begin{tabular}{lcccc}
			\toprule
			Variable & SPX baseline & SPX 3ETF & RUT baseline & RUT 3ETF \\
			\midrule
			Intercept
			& 23.134*** & 19.551*** & 24.577*** & 13.945*** \\
			& (5.713) & (4.434) & (5.407) & (3.805) \\
			
			\(GBM^{\mathrm{OIS},1Y}\)
			& -0.548*** & -0.128*** & -0.555*** & -0.109*** \\
			& (0.170) & (0.047) & (0.124) & (0.031) \\
			
			\(GBM^{\mathrm{OIS},10Y}\)
			& 0.411** & -- & 0.541*** & -- \\
			& (0.172) & -- & (0.130) & -- \\
			
			\(GBM^{\mathrm{IEFA},70}\)
			& -- & -0.00838*** & -- & -0.00749*** \\
			& -- & (0.00146) & -- & (0.00111) \\
			
			\(GBM^{\mathrm{IGOV},441}\)
			& -- & 0.0401** & -- & 0.00830 \\
			& -- & (0.0160) & -- & (0.0130) \\
			
			\(GBM^{\mathrm{IAU},315}\)
			& -- & 0.0142** & -- & 0.0232*** \\
			& -- & (0.00657) & -- & (0.00486) \\
			
			\(BA^{med}/\tau\)
			& 0.256*** & 0.195*** & 0.130*** & 0.137*** \\
			& (0.0635) & (0.0673) & (0.0225) & (0.0256) \\
			
			\(NFCI\)
			& -25.839** & -34.798*** & -23.961** & -47.980*** \\
			& (10.359) & (7.712) & (10.013) & (6.733) \\
			\midrule
			\(R^2\)
			& 0.3124 & 0.4023 & 0.2809 & 0.3613 \\
			Adj.\ \(R^2\)
			& 0.3123 & 0.4022 & 0.2807 & 0.3611 \\
			RMSE (bp)
			& 13.199 & 12.306 & 13.951 & 13.148 \\
			MAE (bp)
			& 8.682 & 7.740 & 10.103 & 9.417 \\
			\bottomrule
		\end{tabular}
		\begin{flushleft}
			\footnotesize
			Notes: Parentheses report date-based HAC (Newey--West) standard errors with a 21-trading-day lag.
			***, **, and * denote significance at the 1\%, 5\%, and 10\% levels.
			The baseline includes both OIS 1Y and OIS 10Y GBM terms.
			The 3ETF specification drops the OIS 10Y term and includes GBM terms based on IEFA, IGOV, and IAU.
		\end{flushleft}
	\end{table}
	\FloatBarrier
	
	Table~\ref{tab:coef_is_compare} shows that the 3ETF specification changes the
	composition of the opportunity-cost block without eliminating the OIS channel.
	In the baseline, the OIS 1Y term is negative and the OIS 10Y term is positive in
	both markets. In the 3ETF specification, OIS 10Y is removed, but OIS 1Y remains
	negative and statistically significant. Thus, the short-horizon OIS component
	survives the asset extension; what changes is the representation of the slower
	opportunity-cost component.
	
	The asset coefficients have a clear conditional pattern. IEFA enters negatively
	and is strongly significant in both SPX and RUT. Conditional on OIS rates,
	trading frictions, and financial conditions, a stronger developed ex-U.S. equity
	component is associated with a lower carry gap. This is consistent with IEFA
	capturing a global risk-bearing or capital-availability state: when that state
	is stronger, the carry-space wedge is smaller.
	
	IAU enters with the opposite sign. The gold component is positive and
	statistically significant in both markets, suggesting that safe-haven or
	risk-off states are associated with a wider carry gap. IGOV is also positive,
	but its precision is market-dependent: it is significant in SPX and weaker in
	RUT. I therefore interpret IGOV as a slower foreign-bond and duration-related
	component, rather than as the most robust element of the 3ETF block.
	
	The controls retain stable signs. The bid--ask measure is positive in all
	specifications, consistent with wider option-market trading frictions being
	associated with a larger carry-space wedge. NFCI is negative in all
	specifications. Because NFCI is estimated jointly with OIS terms and
	low-frequency asset components, I treat this sign as a conditional control
	relationship rather than as a standalone financial-conditions channel.
	
	\FloatBarrier
	\begin{table}[H]
		\centering
		\onehalfspacing
		\footnotesize
		\caption{Stability of LOYO training coefficients: 3ETF specification}
		\label{tab:loyo_coef_stability_3etf}
		\begin{tabular}{lllrrr}
			\toprule
			Market & Specification & Variable & Mean coefficient & Min--Max & Sign frequency \\
			\midrule
			SPX & 3ETF & \(GBM^{\mathrm{OIS},1Y}\)
			& -0.133 & [-0.254, -0.093] & \(+0/-10\) \\
			SPX & 3ETF & \(GBM^{\mathrm{IEFA},70}\)
			& -0.00840 & [-0.01051, -0.00719] & \(+0/-10\) \\
			SPX & 3ETF & \(GBM^{\mathrm{IGOV},441}\)
			& 0.03928 & [0.01142, 0.05924] & \(+10/-0\) \\
			SPX & 3ETF & \(GBM^{\mathrm{IAU},315}\)
			& 0.01527 & [0.00831, 0.02916] & \(+10/-0\) \\
			\midrule
			RUT & 3ETF & \(GBM^{\mathrm{OIS},1Y}\)
			& -0.106 & [-0.187, -0.066] & \(+0/-10\) \\
			RUT & 3ETF & \(GBM^{\mathrm{IEFA},70}\)
			& -0.00746 & [-0.00950, -0.00624] & \(+0/-10\) \\
			RUT & 3ETF & \(GBM^{\mathrm{IGOV},441}\)
			& 0.00924 & [-0.02026, 0.03918] & \(+9/-1\) \\
			RUT & 3ETF & \(GBM^{\mathrm{IAU},315}\)
			& 0.02316 & [0.01610, 0.03564] & \(+10/-0\) \\
			\bottomrule
		\end{tabular}
	\end{table}
	\FloatBarrier
	
	Table~\ref{tab:loyo_coef_stability_3etf} shows that these signs are not driven
	by a single full-sample estimate. Across LOYO training folds, OIS 1Y and IEFA
	are negative in all 10 folds for both markets. IAU is positive in all 10 folds
	for both markets. IGOV is positive in all SPX folds and in 9 of 10 RUT folds.
	The coefficient structure used for out-of-sample evaluation is therefore broadly
	stable.
	
	Overall, the coefficient evidence supports the same conclusion as the fit
	statistics. The carry gap is not fully summarized by the OIS-only path-risk
	block. The short OIS component remains important, but low-frequency
	outside-option components, especially developed ex-U.S. equity and gold, explain
	a stable residual part of the carry gap.
	
	\section{Drift versus Outside Options}
	\label{sec:horse_race}
	
	This section connects the 3ETF result to the drift-extended specification of
	\citet{Shin26b}. That paper shows that an own-index drift proxy,
	\(\hat{\mu}\), improves the OIS-only carry-gap specification through an
	\(OIS1Y\times\hat{\mu}^{504}\times\tau\) term. The question here is narrower:
	does this drift proxy retain independent explanatory power once the broader
	outside-option block is included?
	
	The exercise is an absorption test, not a rejection test of the drift channel.
	Because the true physical drift \(\mu\) is unobserved, \(\hat{\mu}^{504}\) is an
	empirical rolling-trend proxy. If it adds little after IEFA, IGOV, and IAU are
	included, the natural interpretation is not that drift is irrelevant. Rather,
	the own-index \(\hat{\mu}\) proxy is likely a scalar projection of a broader
	physical-measure opportunity-cost state.
	
	\subsection{Incremental Fit}
	\label{subsec:horse_race_fit}
	
	I compare the 3ETF specification with the 3ETF+drift specification. The 3ETF
	model includes OIS 1Y, IEFA, IGOV, IAU, \(BA^{med}/\tau\), and NFCI. The
	3ETF+drift model adds \(OIS1Y\times\hat{\mu}^{504}\times\tau\).
	
	\FloatBarrier
	\begin{table}[H]
		\centering
		\onehalfspacing
		\footnotesize
		\caption{Incremental performance of the drift proxy after the 3ETF block}
		\label{tab:horse_race_fit}
		\begin{tabular}{lrrrrrrr}
			\toprule
			& \multicolumn{3}{c}{In sample} & \multicolumn{4}{c}{LOYO out of sample} \\
			\cmidrule(lr){2-4}\cmidrule(lr){5-8}
			Market
			& \(R^2\) 3ETF & \(R^2\) +drift & \(\Delta R^2\)
			& Pooled \(R^2\) 3ETF & Pooled \(R^2\) +drift
			& \(\Delta R^2\) & \(\Delta\) mean RMSE \\
			\midrule
			SPX & 0.402 & 0.404 & 0.002 & 0.379 & 0.379 &  0.000 & -0.02 \\
			RUT & 0.361 & 0.361 & 0.000 & 0.318 & 0.300 & -0.018 &  0.15 \\
			\bottomrule
		\end{tabular}
	\end{table}
	\FloatBarrier
	
	Table~\ref{tab:horse_race_fit} shows limited incremental fit from adding the
	drift proxy after the 3ETF block. In sample, SPX gains only 0.002 in \(R^2\),
	while RUT is essentially unchanged. Out of sample, SPX pooled \(R^2\) is
	unchanged at 0.379, whereas RUT pooled \(R^2\) falls from 0.318 to 0.300. Mean
	RMSE improves only marginally in SPX and deteriorates in RUT.
	
	Thus, the drift proxy does not provide a meaningful second layer of explanatory
	power once the outside-option block is already present. The evidence is
	consistent with the drift proxy containing useful low-frequency information in
	the OIS-only specification, but little independent information after the broader
	state vector is included.
	
	\subsection{Drift Loading}
	\label{subsec:horse_race_loading}
	
	The coefficient evidence leads to the same conclusion. Table~\ref{tab:horse_race_loading}
	reports the HAC drift loading in the full sample and the corresponding LOYO
	training-coefficient stability.
	
	\FloatBarrier
	\begin{table}[H]
		\centering
		\onehalfspacing
		\footnotesize
		\caption{Drift loading after the 3ETF block}
		\label{tab:horse_race_loading}
		\begin{tabular}{lrrrr}
			\toprule
			Market
			& HAC coefficient
			& HAC s.e.
			& LOYO mean coefficient
			& LOYO sign frequency \\
			\midrule
			SPX &  0.027 & 0.019 &  0.025 & \(+10/-0\) \\
			RUT & -0.005 & 0.016 & -0.010 & \(+2/-8\) \\
			\bottomrule
		\end{tabular}
		\begin{flushleft}
			\footnotesize
			Notes: The drift term is \(OIS1Y\times\hat{\mu}^{504}\times\tau\).
			HAC standard errors use a 21-trading-day Newey--West lag.
			The LOYO sign frequency reports the sign of the training coefficient across
			the ten holdout folds.
		\end{flushleft}
	\end{table}
	\FloatBarrier
	
	The drift loading is statistically insignificant in both markets. In SPX, the
	coefficient is positive but small. It is also positive in all LOYO folds, so the
	direction does not disappear completely. However, the magnitude is too small to
	generate meaningful incremental fit after the 3ETF block is included. In RUT,
	the full-sample coefficient is slightly negative, and the LOYO coefficient is
	negative in 8 out of 10 folds. The drift proxy therefore does not behave as a
	stable additional state variable in RUT.
	
	By contrast, the core 3ETF structure remains stable after the drift term is
	added. IEFA remains negative in both markets, IAU remains positive in both
	markets, and the OIS 1Y term remains negative. The stable structure is therefore
	not the additional own-index drift term, but the outside-option block layered on
	top of the short-horizon OIS component.
	
	These results extend rather than overturn \citet{Shin26b}. The earlier drift
	evidence shows that the carry gap is not fully explained by OIS-based path risk,
	trading frictions, and financial conditions. The present horse race shows that
	the own-index \(\hat{\mu}\) proxy is not the final state variable. It is better
	interpreted as a scalar projection of a broader physical-measure
	investment-opportunity state. Once that state is represented by IEFA, IGOV, and
	IAU, little independent information remains in the own-index drift proxy.
	
	\section{Robustness}
	\label{sec:robust}
	
	This section tests four threats to the main 3ETF result: broad-dollar exposure,
	common-factor collapse, lookback-horizon selection, and ETF-block choice. First,
	I remove a broad U.S.-dollar component from the IEFA, IGOV, and IAU price
	series and repeat the main analysis. Second, I use PCA and residualization to
	ask whether the 3ETF block collapses to a single latent factor. Third, I use a
	nested LOYO procedure to check whether horizon selection leaks information from
	the holdout year. Fourth, I compare the main developed ex-U.S. block with
	U.S.-centered and emerging-market-centered alternatives.
	
	\subsection{Dollar Adjustment}
	\label{subsec:fx_neutral_broad_dollar}
	
	The first test asks whether the 3ETF result is merely a disguised broad-dollar
	cycle. I construct broad-dollar-adjusted ETF price series by subtracting the log
	broad-dollar index, \texttt{DTWEXBGS}, from the log prices of IEFA, IGOV, and
	IAU, and then recompute the rolling OLS slope series from these adjusted price
	paths. Appendix~\ref{app:implementation} describes the calendar alignment and
	previous-observation fill used to match \texttt{DTWEXBGS} to the ETF trading
	calendar.
	
	This adjustment is deliberately mechanical. It is not intended to construct
	perfect currency-hedged ETF returns. It asks the narrower question of whether
	the 3ETF block continues to explain the carry gap after removing a broad
	U.S.-dollar component from the asset price series. As in the main 3ETF
	specification, the broad-dollar-adjusted model keeps \(GBM^{\mathrm{OIS},1Y}\),
	drops \(GBM^{\mathrm{OIS},10Y}\), and adds the adjusted IEFA, IGOV, and IAU GBM
	terms. The OIS-only baseline is unchanged.
	
	\FloatBarrier
	\begin{table}[H]
		\centering
		\onehalfspacing
		\footnotesize
		\caption{In-sample performance after broad-dollar adjustment}
		\label{tab:main_is_compare_fxn}
		\begin{tabular}{lcc}
			\toprule
			Metric & SPX & RUT \\
			\midrule
			Baseline \(R^2\) & 0.312 & 0.281 \\
			Main 3ETF \(R^2\) & 0.402 & 0.361 \\
			FXN 3ETF \(R^2\) & 0.403 & 0.368 \\
			FXN--main \(R^2\) & 0.001 & 0.007 \\
			\(\Delta R^2\) vs. baseline & 0.091 & 0.087 \\
			FXN RMSE (bp) & 12.296 & 13.079 \\
			\bottomrule
		\end{tabular}
	\end{table}
	\FloatBarrier
	
	Table~\ref{tab:main_is_compare_fxn} shows that broad-dollar adjustment does not
	weaken the in-sample result. In SPX, \(R^2\) changes from 0.402 in the main 3ETF
	specification to 0.403 after dollar adjustment. In RUT, it rises from 0.361 to
	0.368. The adjusted 3ETF block therefore preserves essentially the same fit
	gain over the OIS-only baseline.
	
	If the main result were only a broad-dollar factor in disguise, removing the
	broad-dollar component from the ETF price paths should materially reduce
	explanatory power. It does not. Broad-dollar variation may still be part of the
	global state, especially for the unhedged foreign-bond component, but it does
	not account for the aggregate explanatory content of the 3ETF block.
	
	\subsubsection{Out-of-Sample Performance}
	
	\FloatBarrier
	\begin{table}[H]
		\centering
		\onehalfspacing
		\footnotesize
		\caption{LOYO out-of-sample performance: baseline, main 3ETF, and broad-dollar-adjusted 3ETF}
		\label{tab:oos_summary_fxn}
		\begin{tabular}{llrrrrr}
			\toprule
			Market & Specification & Mean \(R^2\) & Median \(R^2\) & Pooled \(R^2\) & Mean RMSE (bp) & Mean correlation \\
			\midrule
			SPX & Baseline  & 0.059 & 0.130 & 0.221 & 13.947 & 0.205 \\
			SPX & Main 3ETF & 0.288 & 0.195 & 0.379 & 12.369 & 0.373 \\
			SPX & FXN 3ETF  & 0.283 & 0.212 & 0.373 & 12.461 & 0.399 \\
			\midrule
			RUT & Baseline  & 0.075 & 0.108 & 0.171 & 15.075 & 0.243 \\
			RUT & Main 3ETF & 0.237 & 0.211 & 0.318 & 13.825 & 0.356 \\
			RUT & FXN 3ETF  & 0.251 & 0.222 & 0.320 & 13.786 & 0.369 \\
			\bottomrule
		\end{tabular}
	\end{table}
	\FloatBarrier
	
	Table~\ref{tab:oos_summary_fxn} shows that the broad-dollar-adjusted 3ETF
	block also survives LOYO validation. In SPX, pooled OOS \(R^2\) changes only
	from 0.379 in the main 3ETF model to 0.373 after dollar adjustment. In RUT, it
	changes from 0.318 to 0.320. Mean \(R^2\), median \(R^2\), RMSE, and
	correlation are also close to the main 3ETF results.
	
	Thus, the OOS result is not simply a comparison against the OIS-only baseline.
	The dollar-adjusted 3ETF block performs almost as well as the original 3ETF
	block. This is the key evidence against the interpretation that the main result
	is just an unremoved broad-dollar cycle.
	
	\FloatBarrier
	\begin{table}[H]
		\centering
		\onehalfspacing
		\footnotesize
		\caption{Year-by-year LOYO out-of-sample performance: SPX, broad-dollar-adjusted 3ETF}
		\label{tab:oos_yearly_spx_fxn}
		\begin{tabular}{ccccccc}
			\toprule
			Holdout year &
			\multicolumn{3}{c}{Baseline} &
			\multicolumn{3}{c}{FXN 3ETF} \\
			\cmidrule(lr){2-4}\cmidrule(lr){5-7}
			& \(R^2\) & RMSE (bp) & Corr. & \(R^2\) & RMSE (bp) & Corr. \\
			\midrule
			2016 &  0.185 &  8.970 &  0.194 & -0.015 & 10.012 &  0.389 \\
			2017 &  0.074 & 10.069 &  0.226 &  0.274 &  8.918 &  0.409 \\
			2018 &  0.221 & 26.304 &  0.259 &  0.167 & 27.200 &  0.205 \\
			2019 &  0.023 &  8.934 & -0.011 &  0.202 &  8.074 &  0.404 \\
			2020 & -1.221 & 18.423 & -0.294 &  0.061 & 11.980 &  0.409 \\
			2021 &  0.416 &  7.600 &  0.064 &  0.556 &  6.629 &  0.123 \\
			2022 &  0.357 & 16.040 &  0.347 &  0.491 & 14.275 &  0.300 \\
			2023 &  0.571 & 15.725 &  0.438 &  0.695 & 13.267 &  0.513 \\
			2024 & -0.016 & 15.428 &  0.589 &  0.221 & 13.506 &  0.748 \\
			2025 & -0.022 & 11.980 &  0.242 &  0.178 & 10.744 &  0.487 \\
			\midrule
			Mean   &  0.059 & 13.947 & 0.205 &  0.283 & 12.461 & 0.399 \\
			Median &  0.130 &        &       &  0.212 &        &       \\
			Pooled &  0.221 &        &       &  0.373 &        &       \\
			\bottomrule
		\end{tabular}
	\end{table}
	\FloatBarrier
	
	\FloatBarrier
	\begin{table}[H]
		\centering
		\onehalfspacing
		\footnotesize
		\caption{Year-by-year LOYO out-of-sample performance: RUT, broad-dollar-adjusted 3ETF}
		\label{tab:oos_yearly_rut_fxn}
		\begin{tabular}{ccccccc}
			\toprule
			Holdout year &
			\multicolumn{3}{c}{Baseline} &
			\multicolumn{3}{c}{FXN 3ETF} \\
			\cmidrule(lr){2-4}\cmidrule(lr){5-7}
			& \(R^2\) & RMSE (bp) & Corr. & \(R^2\) & RMSE (bp) & Corr. \\
			\midrule
			2016 & -0.350 & 12.787 &  0.069 &  0.183 &  9.952 &  0.300 \\
			2017 & -0.152 & 10.498 &  0.142 &  0.038 &  9.592 &  0.353 \\
			2018 &  0.213 & 18.830 &  0.121 &  0.133 & 19.763 &  0.235 \\
			2019 & -0.041 & 11.088 & -0.007 &  0.066 & 10.504 &  0.459 \\
			2020 & -0.587 & 22.437 &  0.102 &  0.281 & 15.100 &  0.422 \\
			2021 &  0.477 & 12.368 &  0.431 &  0.502 & 12.071 &  0.382 \\
			2022 &  0.308 & 16.619 &  0.368 &  0.265 & 17.130 &  0.221 \\
			2023 &  0.661 & 13.922 &  0.354 &  0.738 & 12.247 &  0.454 \\
			2024 &  0.153 & 14.005 &  0.598 &  0.260 & 13.086 &  0.494 \\
			2025 &  0.064 & 18.193 &  0.248 &  0.041 & 18.411 &  0.368 \\
			\midrule
			Mean   &  0.075 & 15.075 & 0.243 &  0.251 & 13.786 & 0.369 \\
			Median &  0.108 &        &       &  0.222 &        &       \\
			Pooled &  0.171 &        &       &  0.320 &        &       \\
			\bottomrule
		\end{tabular}
	\end{table}
	\FloatBarrier
	
	Tables~\ref{tab:oos_yearly_spx_fxn} and~\ref{tab:oos_yearly_rut_fxn} report the
	year-by-year results. The broad-dollar-adjusted 3ETF specification beats the
	baseline in 8 of 10 SPX holdouts and 7 of 10 RUT holdouts. It also produces
	positive OOS \(R^2\) in 9 of 10 SPX holdouts and in all 10 RUT holdouts,
	compared with 7 and 6 positive holdouts for the baseline.
	
	The 2020 holdout shows that the stress-year repair is not eliminated by dollar
	adjustment. In SPX, the baseline OOS \(R^2\) is \(-1.221\), while the adjusted
	3ETF model gives 0.061. In RUT, the corresponding values are \(-0.587\) and
	0.281. The adjusted asset block therefore still mitigates the largest
	level-calibration failure of the OIS-only baseline.
	
	The result is not uniform across every year. The adjusted specification is
	weaker than the baseline in SPX in 2016 and 2018, and in RUT in 2018, 2022, and
	2025. I therefore interpret the test as evidence of out-of-sample stability
	after dollar adjustment, not as a claim of uniformly superior forecasting
	performance in every holdout year.
	
	\subsubsection{Coefficient Structure}
	
	The coefficient evidence clarifies which parts of the 3ETF block survive dollar
	adjustment. Rather than re-estimating the full economic interpretation of all
	controls, I focus on the adjusted asset terms and the retained OIS 1Y component.
	
	\FloatBarrier
	\begin{table}[H]
		\centering
		\onehalfspacing
		\footnotesize
		\caption{Key coefficients after broad-dollar adjustment}
		\label{tab:coef_summary_fxn}
		\begin{tabular}{lcccc}
			\toprule
			Variable
			& SPX coefficient
			& SPX LOYO signs
			& RUT coefficient
			& RUT LOYO signs \\
			\midrule
			\(GBM^{\mathrm{OIS},1Y}\)
			& \(-0.1735^{***}\) & \(+0/-10\)
			& \(-0.1157^{***}\) & \(+0/-10\) \\
			& \((0.0450)\) & 
			& \((0.0275)\) &  \\
			
			\(GBM^{\mathrm{IEFA,FXN},70}\)
			& \(-0.00749^{***}\) & \(+0/-10\)
			& \(-0.00620^{***}\) & \(+0/-10\) \\
			& \((0.00122)\) &
			& \((0.00097)\) & \\
			
			\(GBM^{\mathrm{IGOV,FXN},441}\)
			& \(0.01264\) & \(+9/-1\)
			& \(0.00252\) & \(+8/-2\) \\
			& \((0.01259)\) &
			& \((0.00983)\) & \\
			
			\(GBM^{\mathrm{IAU,FXN},315}\)
			& \(0.02245^{***}\) & \(+10/-0\)
			& \(0.02425^{***}\) & \(+10/-0\) \\
			& \((0.00603)\) &
			& \((0.00415)\) & \\
			\bottomrule
		\end{tabular}
		\begin{flushleft}
			\footnotesize
			Notes: Parentheses report date-based HAC Newey--West standard errors with a
			21-trading-day lag. ***, **, and * denote significance at the 1\%, 5\%, and
			10\% levels. LOYO signs report the sign frequency of the corresponding
			training coefficient across the ten holdout folds.
		\end{flushleft}
	\end{table}
	\FloatBarrier
	
	Table~\ref{tab:coef_summary_fxn} shows that the dollar-adjusted signal is
	concentrated in IEFA and IAU. The adjusted IEFA term is negative, highly
	significant, and negative in every LOYO fold for both markets. The adjusted IAU
	term is positive, highly significant, and positive in every LOYO fold. These two
	components therefore carry the most stable FX-orthogonal part of the 3ETF block.
	
	IGOV behaves differently. After broad-dollar adjustment, its coefficient remains
	positive but is no longer statistically significant in either market, and its
	LOYO sign stability weakens, especially in RUT. This suggests that the original
	IGOV signal is partly dollar-sensitive. IGOV is still useful as a slow
	foreign-bond state variable in the main specification, but it is not the clean
	FX-orthogonal core of the result.
	
	The OIS 1Y term remains negative and significant in both markets, with stable
	LOYO signs. Dollar adjustment therefore does not eliminate the short-horizon OIS
	path-risk channel. Overall, the test supports the main interpretation: the 3ETF
	result is not reducible to a broad-dollar cycle, but its components are
	heterogeneous. IEFA and IAU provide robust dollar-adjusted channels, while IGOV
	is a weaker dollar-sensitive slow component.
	
	\subsection{Common-Factor Tests}
	\label{subsec:pca_orthogonalization}
	
	This subsection asks whether the 3ETF block is only repeated exposure to one
	latent global factor. If IEFA, IGOV, and IAU were close substitutes for the same
	cycle, the asset-slope series should be highly correlated and the first
	principal component should dominate their variation.
	
	The pairwise correlations are modest: 0.088 between IEFA and IGOV, 0.170
	between IEFA and IAU, and 0.381 between IGOV and IAU. Thus, even before
	rotation, the three slope series do not look like different labels for a single
	common factor.
	
	\FloatBarrier
	\begin{table}[H]
		\centering
		\onehalfspacing
		\footnotesize
		\caption{Principal-component decomposition of the three asset-slope series}
		\label{tab:pca_asset_slopes}
		\begin{tabular}{lrrrr}
			\toprule
			Component & Explained variance & IEFA loading & IGOV loading & IAU loading \\
			\midrule
			PC1 & 48.5\% & 0.375 & 0.637 & 0.674 \\
			PC2 & 31.2\% & 0.913 & -0.382 & -0.146 \\
			PC3 & 20.3\% & -0.164 & -0.670 & 0.724 \\
			\bottomrule
		\end{tabular}
	\end{table}
	\FloatBarrier
	
	Table~\ref{tab:pca_asset_slopes} confirms that the 3ETF block is not
	one-dimensional. PC1 explains only 48.5\% of total variance and loads mainly on
	IGOV and IAU. PC2 explains another 31.2\% and is strongly IEFA-dominant. The
	foreign-equity component therefore does not collapse into the slower
	IGOV--IAU component. PC3 mainly separates IGOV from IAU.
	
	I also replace the raw asset GBM terms with three orthogonal principal-component
	GBM terms. Because the three PCs preserve the linear span of the original asset
	block, the in-sample fit is unchanged: \(R^2=0.402\) in SPX and \(R^2=0.361\)
	in RUT, the same as in the raw 3ETF specification. The point is not that PCA
	raises fit, but that the fit is not an artifact of collinearity among the raw
	ETF variables. The explanatory power belongs to the low-frequency state space
	spanned by the assets.
	
	Finally, I residualize each asset slope on the other two. The first-stage
	\(R^2\) values are 0.029 for IEFA, 0.146 for IGOV, and 0.164 for IAU. IEFA is
	therefore almost orthogonal to the other two slopes, and even IGOV and IAU retain
	substantial asset-specific variation. The residualized components continue to
	explain the carry gap, indicating that asset-specific variation outside the
	other two slopes still contains alignment information.
	
	Overall, these tests weaken the single-common-factor interpretation. The 3ETF
	block contains a slow IGOV--IAU component, a distinct IEFA-dominant component,
	and residual variation separating IGOV from IAU. It is better interpreted as a
	multi-channel outside-option structure with heterogeneous frequencies, not as
	three noisy versions of one latent global factor.
	
	\subsection{Nested Horizon Selection}
	\label{subsec:nested_horizon_selection}
	
	The main specification fixes the asset lookback horizons at 70 trading days for
	IEFA, 441 days for IGOV, and 315 days for IAU. Because these horizons are
	selected from asset-by-asset horizon profiles, full-sample horizon choice could
	overstate out-of-sample performance. I address this concern with a nested LOYO
	procedure.
	
	For each holdout year, I first remove that year from the sample. I then select
	the IEFA, IGOV, and IAU lookback windows using only the remaining training
	years, with the equal-weighted average of SPX and RUT in-sample \(R^2\) as the
	objective. The selected horizons are then fixed, coefficients are estimated on
	the same training years, and performance is evaluated in the excluded holdout
	year. Thus, the holdout year is used neither for coefficient estimation nor for
	horizon selection.
	
	\FloatBarrier
	\begin{table}[H]
		\centering
		\onehalfspacing
		\footnotesize
		\caption{Nested horizon selection: distribution of selected lookback windows}
		\label{tab:nested_horizon_distribution}
		\begin{tabular}{lrrrrrr}
			\toprule
			Asset & Mean & Median & Mode & Min & Max & Unique values \\
			\midrule
			IEFA & 78.1 & 79.0 & 79 & 65 & 87 & 5 \\
			IGOV & 317.8 & 319.0 & 318 & 264 & 366 & 7 \\
			IAU  & 325.6 & 328.5 & 322 & 309 & 335 & 8 \\
			\bottomrule
		\end{tabular}
	\end{table}
	\FloatBarrier
	
	Table~\ref{tab:nested_horizon_distribution} shows that the nested search does
	not mechanically reproduce the main horizons. The selected IGOV horizon is
	shorter than the 441-day main specification horizon. However, the qualitative
	frequency ordering remains intact: IEFA is repeatedly selected at a short
	horizon, while IGOV and IAU are selected at slower horizons. Applying the same
	search to the full sample selects IEFA = 79, IGOV = 319, and IAU = 324. No
	fold-level solution lies on the search boundary.
	
	\FloatBarrier
	\begin{table}[H]
		\centering
		\onehalfspacing
		\footnotesize
		\caption{Nested horizon-selection LOYO performance}
		\label{tab:nested_summary}
		\begin{tabular}{llrrr}
			\toprule
			Market/score & Statistic & Baseline & 3ETF & Difference/count \\
			\midrule
			SPX
			& Mean \(R^2\)   & 0.059 & 0.281 & 0.222 \\
			& Median \(R^2\) & 0.130 & 0.197 & 0.067 \\
			& Positive \(\Delta R^2\) years & -- & -- & 8/10 \\
			\midrule
			RUT
			& Mean \(R^2\)   & 0.075 & 0.212 & 0.137 \\
			& Median \(R^2\) & 0.108 & 0.134 & 0.026 \\
			& Positive \(\Delta R^2\) years & -- & -- & 6/10 \\
			\midrule
			Equal-weight
			& Mean \(R^2\)   & 0.067 & 0.246 & 0.180 \\
			& Median \(R^2\) & 0.045 & 0.163 & 0.118 \\
			& Positive \(\Delta R^2\) years & -- & -- & 8/10 \\
			\bottomrule
		\end{tabular}
	\end{table}
	\FloatBarrier
	
	Table~\ref{tab:nested_summary} shows that the asset-return extension continues
	to improve LOYO performance even when horizons are selected only inside the
	training sample. Mean OOS \(R^2\) rises from 0.059 to 0.281 in SPX and from
	0.075 to 0.212 in RUT. The equal-weighted score rises from 0.067 to 0.246. The
	gain is positive in 8 out of 10 SPX holdouts, 6 out of 10 RUT holdouts, and
	8 out of 10 equal-weighted holdouts.
	
	The gains are smaller than in the fixed-horizon main specification, which is
	expected because the nested design removes full-sample horizon information. But
	the improvement over the OIS-only baseline remains. Using yearly RMSE and
	holdout sample size to compute pooled OOS \(R^2\), the nested score rises from
	0.221 to 0.336 in SPX and from 0.171 to 0.287 in RUT. In the 2020 stress
	holdout, the nested 3ETF specification also repairs much of the baseline
	failure: OOS \(R^2\) improves from \(-1.221\) to 0.054 in SPX and from
	\(-0.587\) to 0.256 in RUT.
	
	Overall, the nested exercise supports the main conclusion. Even when lookback
	windows are selected using training-sample information only, the data again
	select a fast foreign-equity component and slower foreign-bond and gold
	components, and the 3ETF block retains higher out-of-sample explanatory power
	than the OIS-only baseline. The main result is therefore difficult to explain as
	an artifact of ex post full-sample horizon selection alone.
	
	\subsection{Alternative Blocks}
	\label{subsec:alternative_blocks}
	
	Finally, I test whether the main result depends on the specific IEFA--IGOV--IAU
	combination. I compare the main developed ex-U.S. block with two alternatives:
	a U.S.-centered block, VTI(42), BND(252), and IAU(300), and an
	emerging-market-centered block, IEMG(63), EBND(126), and IAU(300). The lookback
	windows for the alternative blocks are selected through separate horizon scans.
	As in the main 3ETF specification, each asset-augmented model keeps
	\(GBM^{\mathrm{OIS},1Y}\) and excludes \(GBM^{\mathrm{OIS},10Y}\).
	
	\FloatBarrier
	\begin{table}[H]
		\centering
		\onehalfspacing
		\footnotesize
		\caption{Performance comparison across alternative 3ETF asset blocks}
		\label{tab:alt_asset_combo_compare}
		\begin{tabular}{llccc}
			\toprule
			Specification & Market & IS \(R^2\) & Mean OOS \(R^2\) & Pooled OOS \(R^2\) \\
			\midrule
			Baseline & SPX & 0.312 & 0.059 & 0.221 \\
			& RUT & 0.281 & 0.075 & 0.171 \\
			\midrule
			Main 3ETF & SPX & 0.402 & 0.288 & 0.379 \\
			& RUT & 0.361 & 0.237 & 0.318 \\
			\midrule
			U.S.-centered & SPX & 0.364 & 0.253 & 0.326 \\
			& RUT & 0.339 & 0.209 & 0.270 \\
			\midrule
			Emerging-market & SPX & 0.385 & 0.243 & 0.331 \\
			& RUT & 0.362 & 0.262 & 0.342 \\
			\bottomrule
		\end{tabular}
		\begin{flushleft}
			\footnotesize
			Notes: Main 3ETF = IEFA(70) + IGOV(441) + IAU(315);
			U.S.-centered = VTI(42) + BND(252) + IAU(300);
			Emerging-market = IEMG(63) + EBND(126) + IAU(300).
			Out-of-sample performance is evaluated using LOYO validation.
			All asset-augmented specifications keep \(GBM^{\mathrm{OIS},1Y}\)
			and exclude \(GBM^{\mathrm{OIS},10Y}\).
		\end{flushleft}
	\end{table}
	\FloatBarrier
	
	Table~\ref{tab:alt_asset_combo_compare} shows that the asset-extension result
	does not depend on a single ETF combination. All three asset-augmented blocks
	improve on the OIS-only baseline in both markets, both in sample and out of
	sample. The main developed ex-U.S. block gives the strongest SPX performance,
	with pooled OOS \(R^2=0.379\). In RUT, the emerging-market block performs best
	out of sample, with pooled OOS \(R^2=0.342\).
	
	The weaker performance of the U.S.-centered block should be interpreted
	conditionally. The baseline already contains substantial U.S.-centered
	information through OIS rates, VIX/RVX, bid--ask frictions, and NFCI. The
	foreign-tilted blocks therefore appear to contain more residual outside-option
	information after the baseline controls are included.
	
	Overall, the alternative-block evidence supports the main interpretation. The
	IEFA--IGOV--IAU block is the most balanced specification, especially for SPX,
	but it is not a uniquely lucky ETF combination. Other asset-allocation blocks
	also improve on the baseline, which suggests that the carry gap is aligned with
	residual outside-option states more generally.
	
	\section{Discussion}
	\label{sec:discuss}
	
	\subsection{A Conditional Outside-Option Interpretation}
	\label{subsec:discuss_conditional}
	
	The 3ETF result should be interpreted conditionally, not as a claim that foreign
	assets are unconditionally better outside-option proxies than U.S. assets. The
	baseline is already strongly U.S.-centered. OIS rates capture the funding
	component, VIX and RVX enter the GBM scale as high-frequency market-specific
	volatility states, \(BA^{med}/\tau\) captures option-market trading frictions,
	and NFCI partly absorbs slow U.S. financial-condition and risk-bearing
	variation. The asset-extension exercise therefore asks which asset-return
	components contain incremental information after this baseline state vector has
	already been controlled for.
	
	Equivalently, the relevant object is not the unconditional information in an
	asset-return component \(Z_a\), but its residual information after conditioning
	on the baseline:
	\[
	\Delta R_a^2
	=
	R^2(CG \mid X_{\mathrm{base}}, Z_a)
	-
	R^2(CG \mid X_{\mathrm{base}}).
	\]
	The empirical result is therefore not that U.S. assets are irrelevant. Rather,
	much of the U.S.-centered state space is already represented by the baseline
	controls, leaving less residual room for U.S. asset returns to improve the
	regression.
	
	This interpretation helps explain why the developed ex-U.S. block performs
	better than the U.S.-centered block. VTI and BND may be valid outside options in
	economic terms, but their incremental contribution is evaluated after OIS rates,
	VIX/RVX, NFCI, and option-market frictions have already been partialed out.
	IEFA, IGOV, and IAU appear to contain residual global information less directly
	spanned by this U.S.-centered baseline.
	
	The components of the 3ETF block should therefore be read as conditional
	residual channels rather than structural pricing factors. IEFA captures a global
	risk-bearing component not fully summarized by domestic volatility and financial
	conditions. IAU captures a safe-haven or liquidity-demand component. IGOV is a
	more conditional slow component, combining foreign duration, safe-asset demand,
	and dollar-sensitive global funding variation. The point is not that these
	assets directly price U.S. index options, but that they proxy for residual
	outside-option states faced by finite-capital parity enforcers.
	
	This interpretation is related in spirit to the ICAPM logic of
	\citet{Merton73}. Finite-capital arbitrageurs care not only about the current
	funding rate, but also about the investment-opportunity set available while the
	parity trade is being financed and margined. The exercise here is more modest
	than an ICAPM estimation: I do not recover a structural stochastic discount
	factor or identify primitive state variables. The ETF terms are reduced-form
	proxies for residual investment-opportunity states after conditioning on the
	U.S.-centered baseline.
	
	\subsection{Drift, NFCI, Dollar Exposure, and Common Factors}
	\label{subsec:discuss_evidence}
	
	The drift evidence supports the same partial-out interpretation. In
	\citet{Shin26b}, adding the own-index drift proxy \(\hat{\mu}^{504}\) to the
	OIS-only baseline reduces the absolute NFCI loading in both markets. In SPX, the
	NFCI coefficient moves from \(-25.839\) to \(-19.577\). In RUT, it moves from
	\(-23.961\) to \(-22.286\). By contrast, the \(BA^{med}/\tau\) coefficient is
	largely unchanged. This pattern suggests that the drift proxy absorbs part of
	the slow U.S. financial-condition state rather than the option-market
	trading-friction channel.
	
	The 3ETF horse race extends this logic. The own-index drift proxy improves the
	OIS-only specification, but it has little independent explanatory power after
	IEFA, IGOV, and IAU are included. This does not reject the drift channel. It
	suggests that \(\hat{\mu}^{504}\) is a scalar projection of a broader
	physical-measure investment-opportunity state. The 3ETF block captures that
	state more flexibly, leaving less residual role for the single own-index drift
	term.
	
	The broad-dollar adjustment and common-factor tests further discipline this
	interpretation. If the 3ETF result were merely a disguised broad-dollar cycle,
	removing a broad U.S.-dollar component from the ETF price paths should have
	produced a large deterioration. It does not. The adjusted 3ETF block performs
	close to the original 3ETF specification. At the coefficient level, however,
	IGOV weakens after dollar adjustment, indicating that its original signal is
	partly dollar-sensitive.
	
	The PCA and residualization tests also reject the view that the 3ETF block is
	only repeated exposure to one latent factor. The first principal component does
	not dominate the three asset-slope series, the IEFA-dominant component remains
	distinct, and residualized asset components continue to explain the carry gap.
	The evidence is therefore better read as a small multi-channel outside-option
	state space, not as one hidden global factor.
	
	\subsection{P--Q Alignment and Limits}
	\label{subsec:discuss_pq_limits}
	
	The results should be interpreted as reduced-form P--Q alignment, not as
	structural P-to-Q recovery. Put--call parity remains a terminal-payoff identity.
	The physical-measure asset components do not enter the payoff identity, and the
	results do not imply a failure of risk-neutral pricing.
	
	The mechanism is implementation-based. The carry gap is a \(Q\)-measure
	carry-space object extracted from option-implied discounting. But enforcing
	put--call parity requires capital, margin support, funding capacity, and
	balance-sheet usage before maturity. The arbitrageur who supplies this
	enforcement capital faces physical-measure outside options. Thus, even though
	terminal parity is risk-neutral, the implementation premium embedded in the
	carry gap can be aligned with physical-measure investment-opportunity states.
	
	Several limitations remain. The regressions are reduced-form and should not be
	interpreted as causal estimates, a complete stochastic discount factor, or a
	structural model of intermediary optimization. The selected ETFs are proxies,
	not primitive state variables. The lookback horizons and alternative-block
	tests discipline the specification, but they do not identify a unique structural
	state vector.
	
	The main takeaway is conditional. The carry gap is not fully explained by
	OIS-only funding variables, and it is not fully summarized by the own-index
	drift proxy. After U.S.-centered funding, volatility, trading-friction, and
	financial-condition controls are included, the remaining outside-option
	information is more visible in global and ex-U.S. asset components. The 3ETF
	specification is therefore best understood as a parsimonious proxy for residual
	outside-option costs of finite arbitrage capital, rather than as a structural
	claim that foreign assets price U.S. options.
	
	\section{Conclusion}
	\label{sec:conclusion}
	
	This paper tests whether the carry gap in put--call parity is fully explained
	by OIS-based discounting, volatility, trading-friction, and financial-condition
	variables, or whether it also contains residual information aligned with outside
	investment opportunities faced by finite arbitrage capital. The evidence
	supports the latter view. Adding the 3ETF block of IEFA, IGOV, and IAU improves
	the OIS-based baseline in both SPX and RUT options, and the improvement survives
	leave-one-year-out validation and a range of robustness checks, including
	broad-dollar adjustment, common-factor tests, nested horizon selection, and
	alternative asset-block specifications.
	
	The economic interpretation is conditional. The result should not be read as a
	claim that foreign assets are unconditionally better outside-option proxies than
	U.S. assets. The baseline already contains substantial U.S.-centered information
	through OIS rates, VIX/RVX, bid--ask frictions, and NFCI. The 3ETF block is
	therefore informative because it captures residual global investment-opportunity
	states after domestic funding, volatility, trading-friction, and
	financial-condition variables have already been controlled for. In this sense,
	IEFA, IGOV, and IAU are not structural pricing factors, but reduced-form proxies
	for marginal outside-option information left outside the baseline state vector.
	
	The drift evidence is complementary to this interpretation. The own-index drift
	proxy \(\hat{\mu}\) improves the OIS-only baseline, but contributes little after
	the 3ETF block is included. This pattern suggests that \(\hat{\mu}\) is not a
	competing explanation, but a scalar projection of a broader physical-measure
	investment-opportunity state. The 3ETF block provides a more flexible
	representation of that state, while the robustness tests show that its
	explanatory power is not reducible to a broad-dollar factor or to a single
	latent common component.
	
	The results should be interpreted as reduced-form \(P\)--\(Q\) alignment, not as
	structural \(P\)-to-\(Q\) recovery. Put--call parity remains a terminal
	no-arbitrage identity, and the physical-measure asset components do not enter
	option payoffs directly. The point is that enforcing the parity relation in
	actual markets is path-dependent and capital-using. Before maturity,
	arbitrageurs face variation margin, funding needs, interim losses, liquidation
	risk, and alternative uses of capital. The central contribution is therefore to
	show that a nearly closed no-arbitrage relation can still leave a carry-space
	trace of the opportunity cost of finite arbitrage capital. The carry gap is a
	risk-neutral parity object in construction, but its enforcement can reflect
	physical-measure investment opportunities.
	
	\subsection*{Funding}
	This research did not receive any specific grant from funding agencies in the public, commercial, or not-for-profit sectors.
	
	\subsection*{Declaration of AI usage in manuscript preparation}
	During the preparation of this manuscript, the author used ChatGPT (OpenAI) and Claude (Anthropic) for language refinement and structural clarity.
	All outputs were reviewed and edited by the author, who takes full responsibility for the content.
	
	\subsection*{Declaration of interest}
	The author declares no competing interests.
	
	\newpage
	\onehalfspacing

	\onehalfspacing
	\newpage
	\begin{appendices}
		
		\section{Data and Methodological Details}
		\label{app:implementation}
		
		This appendix summarizes the implementation details behind the carry-gap
		construction, asset-return variables, drift proxy, and in-sample and
		out-of-sample evaluation. The main text focuses on economic interpretation and
		model comparison; this appendix records the data processing and estimation
		steps needed for replication.
		
		\subsection{Data and Analysis Sample}
		
		The option data are minute-level NBBO quotes from ThetaData. The analysis uses
		the period from January 4, 2016 to October 31, 2025, which is the common sample
		over which option-market information and the OIS discount curve are both
		available. All main results are computed on this common sample.
		
		SPX and RUT options are European-style index options. I therefore do not adjust
		the put--call-parity construction for early-exercise premia. The empirical
		pipeline is implemented in MATLAB R2025b.
		
		\subsection{Option-Implied Discount Factors and Carry-Gap Construction}
		
		I identify option-implied discount factors using the synthetic-forward
		procedure of \citet{AB21}. For a European call and put with the same date \(t\),
		maturity \(T\), and strike \(K\), put--call parity can be written as
		\begin{equation}
			C_t(K,T)-P_t(K,T)
			=
			B_t(T)\bigl(F_t(T)-K\bigr),
		\end{equation}
		where \(B_t(T)\) is the market-implied discount factor and \(F_t(T)\) is the
		forward value for maturity \(T\). Define the synthetic forward as
		\begin{equation}
			G_t(K,T)=C_t(K,T)-P_t(K,T).
		\end{equation}
		Then
		\begin{equation}
			F_t(T)=\frac{G_t(K,T)}{B_t(T)}+K.
		\end{equation}
		Within each date--maturity cell, \(B_t(T)\) is identified as the value that
		makes the recovered forward price flat across strikes. Operationally, I use the
		cross-sectional linear relation between synthetic forwards and strikes to
		estimate \(\hat{B}_t(T)\) and \(\hat{F}_t(T)\) jointly.
		
		The benchmark discount factor is constructed from the OIS curve. I bootstrap
		daily OIS data to recover maturity-specific discount factors and zero rates,
		and then interpolate maturity-matched OIS discount factors to option maturities.
		As in equation~\eqref{eq:carry_gap_def} in the main text, the carry gap is the
		annualized wedge between the option-implied discount factor and the OIS discount
		factor. The empirical analysis uses \(CG_{i,t}^{bp}\), measured in basis points.
		
		\subsection{Sample Filters and Date--Maturity Aggregation}
		
		The preprocessing is designed to make the cross-sectional identification of
		\(\hat{B}_t(T)\) stable. I first keep only call--put pairs with the same strike
		and maturity. I then apply rule-based filters that remove quotes with abnormal
		prices, excessive bid--ask spreads, insufficient strike coverage within a
		date--maturity cell, or unstable OIS curve construction. The same filters are
		applied symmetrically to SPX and RUT and are fixed before any regression
		estimation.
		
		For each market, I construct date--maturity cells and estimate the
		option-implied discount factor within each eligible cell. I then form a
		date--maturity panel of carry gaps. When multiple eligible observations remain
		within the same date--maturity cell, I aggregate them using the median. Median
		aggregation reduces sensitivity to transient quote noise, stale quotes, and
		isolated cross-sectional outliers.
		
		\subsection{Asset-Return-Based Low-Frequency Components}
		
		The asset-return extension constructs low-frequency return components from ETF
		prices representing broad asset classes. The candidate ETFs are VTI, IEFA,
		IEMG, BND, SCHP, IGOV, EBND, IAU, VNQ, and VNQI. They represent U.S. equity,
		developed ex-U.S. equity, emerging-market equity, U.S. aggregate bonds, U.S.
		inflation-linked bonds, international sovereign bonds, emerging-market sovereign
		bonds, gold, U.S. REITs, and ex-U.S. REITs. The main 3ETF specification uses
		IEFA, IGOV, and IAU.
		
		For asset \(a\), the \(n\)-day low-frequency component is the OLS slope of the
		log price path over the past \(n\) trading days:
		\begin{equation}
			\log P_{a,t-n+\ell}
			=
			\alpha_{a,t}^{(n)}
			+
			b_{a,t}^{(n)}\ell
			+
			u_{a,t-n+\ell},
			\qquad
			\ell=0,1,\dots,n-1.
			\label{eq:appendix_asset_slope}
		\end{equation}
		The slope uses information available through \(t-1\); the contemporaneous price
		at date \(t\) is not used. This timing convention prevents the asset-return
		component from using information from the same date as the carry-gap
		observation.
		
		The OLS slope is less sensitive to endpoint noise than a simple cumulative
		return. In the asset-based GBM term, \(b_{a,t}^{(n)}\) enters as the rate-like
		opportunity-cost component. Because OIS rates are observed in percentage points,
		the OIS-based GBM term uses \(OIS/100\). By contrast, \(b_{a,t}^{(n)}\) is a
		log-price slope, so no additional \(1/100\) adjustment is applied.
		
		\subsection{Own-Index Drift Proxy}
		
		The drift specification constructs a rolling drift proxy from the own-index
		total-return path. Let \(TR_{i,t}\) denote the total-return index for market
		\(i\in\{\mathrm{SPX},\mathrm{RUT}\}\). The \(n\)-day drift proxy is the slope
		from
		\begin{equation}
			\log TR_{i,t-n+\ell}
			=
			a_{i,t}^{(n)}
			+
			b_{i,t}^{(n)}\ell
			+
			u_{i,t-n+\ell}^{(n)},
			\qquad
			\ell=0,1,\dots,n-1.
		\end{equation}
		The drift comparison in the main text uses \(n=504\) trading days. The daily
		slope is annualized as
		\begin{equation}
			\hat{\mu}_{i,t}^{ann}
			=
			252 b_{i,t}^{(n)}.
		\end{equation}
		Here \(\hat{\mu}_{i,t}^{ann}\) is not an observed expected return. It is a
		reduced-form empirical drift proxy extrapolated from the past total-return path.
		
		The drift-preserving GBM proxy is
		\begin{equation}
			GBM_{i,t}^{\hat{\mu},\mathrm{OIS},1Y}
			=
			10^4
			\cdot
			\frac{OIS1Y_t}{100}
			\cdot
			\hat{\mu}_{i,t}^{ann}
			\cdot
			\tau_{i,t}.
		\end{equation}
		This term does not assume that \(\hat{\mu}\) is the true physical drift
		\(\mu\). It is an empirical \(r\hat{\mu}\tau\) proxy for the theoretical
		\(r\mu\tau\) structure that appears in the support-capital approximation when
		physical drift is preserved.
		
		\subsection{Broad-Dollar Adjustment}
		
		The broad-dollar robustness test examines whether the foreign-asset components
		simply proxy for a common U.S.-dollar cycle. I use the broad-dollar index
		\texttt{DTWEXBGS} to construct mechanically adjusted price paths for IEFA,
		IGOV, and IAU. Specifically, I subtract the log broad-dollar index from each
		ETF log price and then recompute the rolling OLS slopes using the same procedure
		as in equation~\eqref{eq:appendix_asset_slope}.
		
		This adjustment should not be interpreted as constructing fully currency-hedged
		ETF returns. It is a diagnostic that removes a broad U.S.-dollar component from
		the ETF price paths. The \texttt{DTWEXBGS} calendar does not perfectly match the
		ETF trading calendar, so missing values are forward-filled using the most recent
		available observation before the adjusted slopes are computed.
		
		The broad-dollar-adjusted specification follows the main 3ETF specification. It
		keeps \(GBM^{\mathrm{OIS},1Y}\), excludes \(GBM^{\mathrm{OIS},10Y}\), and
		includes GBM terms based on adjusted IEFA, IGOV, and IAU slopes. The OIS-only
		baseline is unchanged and includes both OIS 1Y and OIS 10Y GBM terms.
		
		\subsection{Comparison Specifications}
		
		The main analysis compares four specifications:
		\[
		\begin{array}{llll}
			\text{Baseline:}
			& GBM^{\mathrm{OIS},1Y},
			& GBM^{\mathrm{OIS},10Y},
			& BA^{med}/\tau,\ NFCI, \\
			\text{Drift:}
			& \text{Baseline}
			& +\ GBM^{\hat{\mu},\mathrm{OIS},1Y}, & \\
			\text{3ETF:}
			& GBM^{\mathrm{OIS},1Y},
			& GBM^{\mathrm{IEFA},70},\ GBM^{\mathrm{IGOV},441},\ GBM^{\mathrm{IAU},315},
			& BA^{med}/\tau,\ NFCI, \\
			\text{3ETF+drift:}
			& \text{3ETF}
			& +\ GBM^{\hat{\mu},\mathrm{OIS},1Y}. &
		\end{array}
		\]
		The 3ETF specification excludes \(GBM^{\mathrm{OIS},10Y}\) because the
		long-horizon OIS component strongly overlaps with the low-frequency
		outside-option component associated with IGOV. Including both terms destabilizes
		coefficient interpretation, while dropping OIS 10Y causes only limited loss in
		explanatory power. The preferred specification therefore keeps OIS 1Y as the
		short-horizon risk-free opportunity-cost channel and replaces the long-horizon
		component with the 3ETF outside-option block.
		
		\subsection{Regression Estimation and Out-of-Sample Evaluation}
		
		All main regressions are estimated separately for SPX and RUT. Coefficient
		inference uses date-based HAC Newey--West standard errors with a fixed
		21-trading-day lag. This lag allows for approximately one month of residual
		autocorrelation and is longer than the standard automatic lag rule for the daily
		sample, which selects about 8 trading days.
		
		Out-of-sample performance is evaluated using leave-one-year-out validation.
		Each calendar year is held out once. The model is estimated on the remaining
		years, and fit is evaluated in the excluded year. In-sample performance is
		reported using \(R^2\), adjusted \(R^2\), RMSE, and MAE. LOYO performance is
		reported by holdout year and summarized using mean \(R^2\), median \(R^2\),
		pooled \(R^2\), mean RMSE, and mean correlation.
		
	\end{appendices}
	

\begin{thebibliography}{99}
		
		\bibitem[Stoll(1969)]{Stoll69}
		Stoll, H. R. (1969).
		The Relationship between Put and Call Option Prices.
		\textit{The Journal of Finance}, 24(5), 801--824.
		\url{https://doi.org/10.1111/j.1540-6261.1969.tb01694.x}
		
		\bibitem[Merton(1973)]{Merton73}
		Merton, R. C. (1973).
		An Intertemporal Capital Asset Pricing Model.
		\textit{Econometrica}, 41(5), 867--887.
		\url{https://doi.org/10.2307/1913811}
		
		\bibitem[Gould and Galai(1974)]{GG74}
		Gould, J. P., \& Galai, D. (1974).
		Transaction Costs and the Relationship between Put and Call Prices.
		\textit{Journal of Financial Economics}, 1(2), 105--129.
		\url{https://doi.org/10.1016/0304-405X(74)90001-4}
		
		\bibitem[Ross(1976)]{Ross76}
		Ross, S. A. (1976).
		The arbitrage theory of capital asset pricing.
		\textit{Journal of Economic Theory}, 13(3), 341--360.
		\url{https://doi.org/10.1016/0022-0531(76)90046-6}
		
		\bibitem[Klemkosky and Resnick(1979)]{KR79}
		Klemkosky, R. C., \& Resnick, B. G. (1979).
		Put--Call Parity and Market Efficiency.
		\textit{The Journal of Finance}, 34(5), 1141--1155.
		\url{https://doi.org/10.1111/j.1540-6261.1979.tb00061.x}
		
		\bibitem[Chen et al.(1986)]{CRR86}
		Chen, N. F., Roll, R., \& Ross, S. A. (1986).
		Economic Forces and the Stock Market.
		\textit{Journal of Business}, 59, 383--403.
		\url{https://doi.org/10.1086/296344}
		
		\bibitem[Campbell(1993)]{Campbell93}
		Campbell, J. Y. (1993).
		Intertemporal asset pricing without consumption data.
		\textit{The American Economic Review}, 83(3), 487--512.
		\url{http://www.jstor.org/stable/2117530}
		
		\bibitem[Fama and French(1993)]{FF93}
		Fama, E. F., \& French, K. R. (1993).
		Common risk factors in the returns on stocks and bonds.
		\textit{Journal of Financial Economics}, 33(1), 3--56.
		\url{https://doi.org/10.1016/0304-405X(93)90023-5}
		
		\bibitem[Shleifer and Vishny(1997)]{SV97}
		Shleifer, A., \& Vishny, R. W. (1997).
		The Limits of Arbitrage.
		\textit{The Journal of Finance}, 52(1), 35--55.
		\url{https://doi.org/10.1111/j.1540-6261.1997.tb03807.x}
		
		\bibitem[Ackert and Tian(2001)]{AT01}
		Ackert, L. F., \& Tian, Y. S. (2001).
		Efficiency in Index Options Markets and Trading in Stock Baskets.
		\textit{Journal of Banking \& Finance}, 25(9), 1607--1634.
		\url{https://doi.org/10.1016/S0378-4266(00)00145-X}
		
		\bibitem[Gromb and Vayanos(2002)]{GV02}
		Gromb, D., \& Vayanos, D. (2002).
		Equilibrium and Welfare in Markets with Financially Constrained Arbitrageurs.
		\textit{Journal of Financial Economics}, 66(2--3), 361--407.
		\url{https://doi.org/10.1016/S0304-405X(02)00228-3}
		
		\bibitem[Campbell and Vuolteenaho(2004)]{CV04}
		Campbell, J. Y., \& Vuolteenaho, T. (2004).
		Bad beta, good beta.
		\textit{American Economic Review}, 94(5), 1249--1275.
		\url{https://doi.org/10.1257/0002828043052240}
		
		\bibitem[Ofek et al.(2004)]{ORW04}
		Ofek, E., Richardson, M., \& Whitelaw, R. F. (2004).
		Limited arbitrage and short sales restrictions: evidence from the options markets.
		\textit{Journal of Financial Economics}, 74(2), 305--342.
		\url{https://doi.org/10.1016/j.jfineco.2003.05.008}
		
		\bibitem[Petkova(2006)]{Petkova06}
		Petkova, R. (2006).
		Do the Fama--French factors proxy for innovations in predictive variables?
		\textit{The Journal of Finance}, 61(2), 581--612.
		\url{https://doi.org/10.1111/j.1540-6261.2006.00849.x}
		
		\bibitem[Bollerslev et al.(2009)]{BTZ09}
		Bollerslev, T., Tauchen, G., \& Zhou, H. (2009).
		Expected Stock Returns and Variance Risk Premia.
		\textit{The Review of Financial Studies}, 22(11), 4463--4492.
		\url{https://doi.org/10.1093/rfs/hhp008}
		
		\bibitem[Brunnermeier and Pedersen(2009)]{BP09}
		Brunnermeier, M. K., \& Pedersen, L. H. (2009).
		Market Liquidity and Funding Liquidity.
		\textit{The Review of Financial Studies}, 22(6), 2201--2238.
		\url{https://doi.org/10.1093/rfs/hhn098}
		
		\bibitem[G{\^a}rleanu and Pedersen(2011)]{GP11}
		G{\^a}rleanu, N., \& Pedersen, L. H. (2011).
		Margin-based asset pricing and deviations from the law of one price.
		\textit{The Review of Financial Studies}, 24(6), 1980--2022.
		\url{https://doi.org/10.1093/rfs/hhr027}
		
		\bibitem[Mitchell and Pulvino(2012)]{MP12}
		Mitchell, M., \& Pulvino, T. (2012).
		Arbitrage Crashes and the Speed of Capital.
		\textit{Journal of Financial Economics}, 104(3), 469--490.
		\url{https://doi.org/10.1016/j.jfineco.2011.09.002}
		
		\bibitem[He and Krishnamurthy(2013)]{HK13}
		He, Z., \& Krishnamurthy, A. (2013).
		Intermediary asset pricing.
		\textit{American Economic Review}, 103(2), 732--770.
		\url{https://doi.org/10.1257/aer.103.2.732}
		
		\bibitem[Adrian et al.(2014)]{AEM14}
		Adrian, T., Etula, E., \& Muir, T. (2014).
		Financial intermediaries and the cross-section of asset returns.
		\textit{The Journal of Finance}, 69(6), 2557--2596.
		\url{https://doi.org/10.1111/jofi.12189}
		
		\bibitem[Ross(2015)]{Ross15}
		Ross, S. (2015).
		The Recovery Theorem.
		\textit{The Journal of Finance}, 70(2), 615--648.
		\url{https://doi.org/10.1111/jofi.12092}
		
		\bibitem[He et al.(2017)]{HKM17}
		He, Z., Kelly, B., \& Manela, A. (2017).
		Intermediary asset pricing: New evidence from many asset classes.
		\textit{Journal of Financial Economics}, 126(1), 1--35.
		\url{https://doi.org/10.1016/j.jfineco.2017.08.002}
		
		\bibitem[Martin(2017)]{Martin17}
		Martin, I. (2017).
		What Is the Expected Return on the Market?
		\textit{The Quarterly Journal of Economics}, 132(1), 367--433.
		\url{https://doi.org/10.1093/qje/qjw034}
		
		\bibitem[Du et al.(2018)]{DTV18}
		Du, W., Tepper, A., \& Verdelhan, A. (2018).
		Deviations from Covered Interest Rate Parity.
		\textit{The Journal of Finance}, 73(3), 915--957.
		\url{https://doi.org/10.1111/jofi.12620}
		
		\bibitem[Azzone and Baviera(2021)]{AB21}
		Azzone, M., \& Baviera, R. (2021).
		Synthetic Forwards and Cost of Funding in the Equity Derivative Market.
		\textit{Finance Research Letters}, 41, 101841.
		\url{https://doi.org/10.1016/j.frl.2020.101841}
		
		\bibitem[Muravyev et al.(2025)]{MPP25}
		Muravyev, D., Pearson, N. D., \& Pollet, J. M. (2025).
		Why does options market information predict stock returns?
		\textit{Journal of Financial Economics}, 172, 104153.
		\url{https://doi.org/10.1016/j.jfineco.2025.104153}
		
		\bibitem[Shin(2026a)]{Shin26a}
		Shin, U. (2026).
		The Cost of a Free Lunch.
		\textit{SSRN Working Paper}, No.~6407379
		\url{https://dx.doi.org/10.2139/ssrn.6407379}
		
		\bibitem[Shin(2026b)]{Shin26b}
		Shin, U. (2026).
		The P behind Q.
		\textit{SSRN Working Paper}, No.~6762800
		\url{https://dx.doi.org/10.2139/ssrn.6762800}
		
		\bibitem[FRED(2026a)]{FRED_NFCI}
		Chicago Fed National Financial Conditions Index [NFCI] (2026a),
		retrieved from FRED, Federal Reserve Bank of St.\ Louis, April 3, 2026.
		\url{https://fred.stlouisfed.org/series/NFCI}
		
		\bibitem[FRED(2026b)]{DTWEXBGS}
		Board of Governors of the Federal Reserve System (US) (2026b).
		Nominal Broad U.S. Dollar Index [DTWEXBGS],
		retrieved from FRED, Federal Reserve Bank of St.\ Louis, April 3, 2026.
		\url{https://fred.stlouisfed.org/series/DTWEXBGS}.
		
		\bibitem[ThetaData(2026)]{ThetaData}
		ThetaData (2026).
		Historical SPX and RUT option NBBO data.
		Retrieved April 3, 2026, from
		\url{https://www.thetadata.net}.
		
	\end{thebibliography}
\end{document}